%% file: 01_main.tex
\documentclass[runningheads]{llncs}
\usepackage{amsmath,amssymb}
\usepackage{semantic}
\usepackage{stmaryrd}
\usepackage{mathtools}
\usepackage{bbm}
\usepackage{tabularx}
\usepackage{listings}
\usepackage{xcolor}
\usepackage{amsmath}
\usepackage{amsfonts}
\usepackage{amssymb}
\usepackage{fancyref}
\usepackage[lined,linesnumbered,algoruled]{algorithm2e}
\usepackage{algpseudocode}
\usepackage{fontawesome}
\usepackage{todonotes}
\usepackage[bookmarks=true]{hyperref}
\hypersetup{
 backref=true,
 pagebackref=true,
 hyperindex=true,
 colorlinks=true,
 breaklinks=true,
 linktoc=section,
 bookmarks=true,
 bookmarksopen=true,
 pdftitle={Creating Knowdge Graphs Subsets using Shape Expressions},
 pdfauthor={Jose Emilio Labra Gayo}
}

\usepackage{tikz}
\usepackage{letltxmacro} 
\usepackage{graphicx}
\usepackage{makeidx} 
\makeindex 
\usepackage{multirow}
\usepackage{pgfplots}

\usepackage{kg_macros}
\usepackage{biblatex}
\definecolor{ocre}{RGB}{0,51,170}

\begin{document}

\title{Creating Knowledge Graphs Subsets using Shape Expressions}

\author{Jose Emilio Labra Gayo\inst{1}\orcidID{0000-0001-8907-5348}}

\authorrunning{J.E. Labra}

\institute{University of Oviedo, Oviedo, Spain \\
\email{labra@uniovi.es}
}

{\def\addcontentsline#1#2#3{}\maketitle} 

\begin{abstract}
The initial adoption of knowledge graphs by Google and later by big companies has increased their adoption and popularity. 
In this paper we present a formal model for three different types of knowledge graphs which we call RDF-based graphs, property graphs and wikibase graphs.

In order to increase the quality of Knowledge Graphs, several approaches have appeared to describe and validate their contents.  
 Shape Expressions (ShEx) has been proposed as concise language for RDF validation. 
 We give a brief introduction to ShEx and present two extensions that can also be used to describe and validate property graphs (PShEx) and wikibase graphs (WShEx).

One problem of knowledge graphs is the large amount of data they contain, which jeopardizes their practical application. 
In order to palliate this problem, one approach is to create subsets of those knowledge graphs for some domains. 
We propose the following approaches to generate those subsets:
 Entity-matching, simple matching, ShEx matching, ShEx plus Slurp and ShEx plus Pregel which are based on declaratively defining the subsets by either matching some content or by Shape Expressions.
The last approach is based on a novel validation algorithm for ShEx based on 
 the Pregel algorithm that can handle big data graphs 
 and has been implemented on Apache Spark GraphX.


\end{abstract}

\tableofcontents
\input{paper}


\end{document}

%% file: paper.tex
\pagebreak
\input{02_commands}
\input{10_introduction}

\input{15_preliminaries}
\input{20_KnowledgeGraphs}

\input{30_DescribingKGs}
\input{40_KGSubSets}

\input{60_Results}
\input{70_RelatedWork}

\input{80_conclusions}
\input{100_acknowledgments}

\printbibliography


%% file: 02_commands.tex
\newcommand\vartextvisiblespace[1][.5em]{%
  \makebox[#1]{%
    \kern.07em
    \vrule height.3ex
    \hrulefill
    \vrule height.3ex
    \kern.07em
  }
}

\newcommand{\mi}[1]{\ensuremath{\mathit{#1}}}
\newcommand{\cM}[1]{\ensuremath{\text{#1}}}

\newcommand\myfont[1]{\ensuremath{\mathcal{#1}}}


\newcommand\ShapeLabel{L}

\newcommand\triple[3]{\ensuremath{\langle #1,#2,#3 \rangle}}
\newcommand\quadruple[4]{\ensuremath{\langle #1,#2,#3,#4\rangle}}
\newcommand\Graph{\myfont{G}}
\newcommand{\tcs}[2]{tcs(#1,#2)}

\newcommand\VertSet{\myfont{V}}
\newcommand\NodeSet{\myfont{N}}
\newcommand\EdgeSet{\myfont{E}}
\newcommand\LabelSet{\myfont{L}}
\newcommand\PLabelSet{\myfont{T}}
\newcommand\MsgSet{\myfont{M}}

\newcommand{\ItemSet}{\myfont{Q}}
\newcommand{\PropSet}{\myfont{P}}
\newcommand{\EntitySet}{\myfont{E}}
\newcommand{\ValueSet}{\myfont{V}}
\newcommand{\DataValueSet}{\myfont{D}}
\newcommand{\StmtSet}{\rho}
\newcommand{\SubjectSet}{\myfont{S}}
\newcommand{\ObjectSet}{\myfont{O}}
\newcommand{\Nat}{\mathbb{N}}

\newcommand{\qs}[1]{\ensuremath{#1}}
\newcommand{\condType}[1]{\ensuremath{hasType_{#1}}}

\newcommand{\neighsGraph}[3]{#1^{#2}_{#3}}
\newcommand{\partition}[1]{part(#1)}
\newcommand{\FinSet}[1]{\ensuremath{FinSet(#1)}}

\newcommand{\emptyGraph}{\ensuremath{\emptyset}}
\newcommand{\EmptyGraph}{\ensuremath{\emptyset}}
\newcommand{\addTriple}{\ensuremath{\rtimes}}
\newcommand{\unionGraphs}{\ensuremath{\cup}}
\newcommand\neighs[2]{\ensuremath{neighs(#1,#2)}}
\newcommand\nodes[1]{\ensuremath{nodes(#1)}}
\newcommand{\node}{\mi{n}}
\newcommand{\lbl}{\mi{l}}

\newcommand{\typing}{\tau}
\newcommand{\emptyTyping}{[]}
\newcommand{\hasType}[2]{\ensuremath{#1@#2}}
\newcommand{\hasNoType}[2]{\ensuremath{#1@!#2}}
\newcommand\EmptyTyping{[]}
\newcommand\addType[2]{\ensuremath{#1@#2}}
\newcommand\AddTyping[3]{\ensuremath{#1@#2 : #3}}
\newcommand\RemoveTyping[3]{\ensuremath{#1@!#2 : (#3 \setminus #1@#2)}}
\newcommand\Combine[2]{\ensuremath{#1\uplus#2}}
\newcommand\CombineOr[2]{\ensuremath{#1\parallel#2}}
\newcommand{\HasType}[3]{\ensuremath{#1@#2\in{}#3}}
\newcommand{\HasNoType}[3]{\ensuremath{#1@!#2\in{}#3}}
\newcommand\FailTyping{\ensuremath{\mathbbm{E}}}

\newcommand\arc[2]{\ensuremath{\vartextvisiblespace\xrightarrow{#1}#2}}
\newcommand{\qualified}[4] {\ensuremath{\arc{#1}{#2}\{#3,#4\}}}
\newcommand{\qualifiedcard}[3] {\ensuremath{\arc{#1}{#2}~#3}}

\newcommand{\semantics}[4] {\ensuremath{|[#1|]^{#2,#3,#4}}}
\newcommand{\nodeSelector}[2]{\ensuremath{|[#1|]^{#2}}}
\newcommand{\shape}{\ensuremath{\phi}}
\newcommand{\shapeRef}[1] {\ensuremath{@#1}}
\newcommand{\shapeTrue}{\ensuremath{\top}}
\newcommand{\sAnd}{\ensuremath{\wedge}}
\newcommand{\sNot}{\ensuremath{\lnot}}
\newcommand{\unbounded}{\ensuremath{*}}

\newcommand{\shapesSet}{\operatorname{\mathit{shapes}}}

\newcommand\ShapeN[1]{\ensuremath{\ShapeGen{#1}{n}}}
\newcommand\ShapeO[1]{\ensuremath{\ShapeGen{#1}{o}}}
\newcommand\ShapeGen[2]{\ensuremath{\Triples{}^{#1}_#2}}

\newcommand{\ShapeNStar}{\ensuremath{\Triples^{*}}}
\newcommand{\SE}{\ensuremath{\mathcal{S}_n(E)}}
\newcommand\Shape[1]{\ensuremath{\mathcal{S}_n|[#1|]}}
\newcommand{\Triples}{\ensuremath{\Sigma}}

\newcommand\deriv[2]{\ensuremath{\partial_{#1}(#2)}}
\newcommand\nullable{\ensuremath{\nu}}

\newcommand{\defineSchema}{\ensuremath{\longmapsto}}

\newcommand{\oom}{\ensuremath{\infty}}

\newcommand{\Vs}{\ensuremath{V_s}}
\newcommand{\Vp}{\ensuremath{V_p}}
\newcommand{\Vo}{\ensuremath{V_o}}

\newcommand\SymbolSet{\ensuremath{\Delta}}
\newcommand{\bagw}{\ensuremath{w}}
\newcommand{\emptyBag}{\ensuremath{\epsilon}}
\newcommand{\bag[1]}{\{|#1|\}}
\newcommand{\bagSem}[1]{\ensuremath{|[#1|]}}
\newcommand{\matchRbeOpen}[2]{#1\approx{}#2}
\newcommand{\matchRbeClosed}[2]{#1\approxeq{}#2}
\newcommand{\matchRbe}[2]{#1\approxeq{}#2}

\newcommand{\eachOf}[2]{\ensuremath{#1;#2}}
\newcommand{\oneOf}[2]{\ensuremath{#1\mid{}#2}}
\newcommand{\someOf}[2]{\ensuremath{#1\mid{}#2}}

\newcommand{\tripleConstraint}[4]{\ensuremath{\qualified{#1}{#2}{#3}{#4}}}
\newcommand{\shapes}[1]{\ensuremath{shapes^{#1}}}
\newcommand{\props}[1]{\ensuremath{preds(#1)}} 
\newcommand{\preds}[1]{\ensuremath{preds(#1)}}
\newcommand{\descendants}[1]{\ensuremath{descendants(#1)}}
\newcommand{\descendantsS}[2]{\ensuremath{descendants_{#1}(#2)}}

\newcommand\Pass{\ensuremath{\mathbbm{p}}}
\newcommand\Fail{\ensuremath{\mathbbm{f}}}
\newcommand\code{\textit}

\newcommand{\Dunno}{\ensuremath{\theta}}
\newcommand{\Empty}{\ensuremath{\varnothing}}
\newcommand{\None}{\ensuremath{\varnothing}}
\newcommand\Over{\ensuremath{\mathbbm{e}}}
\newcommand\Open{\ensuremath{\mathbbm{z}}}

\newcommand{\eq}{==}

\newcommand{\emptyRule}{\ensuremath{\varepsilon}}
\newcommand{\matchShape}{\ensuremath{\simeq_s}}
\newcommand{\matchRule}{\ensuremath{\simeq_r}}
\newcommand{\matchName}{\ensuremath{\simeq_n}}
\newcommand{\matchValue}{\ensuremath{\simeq_v}}

\newcommand{\match}{\ensuremath{\simeq}}
\newcommand{\context}{\ensuremath{\Gamma}}
\newcommand{\emptyEr}{\ensuremath{\varepsilon}}
\newcommand{\andEr}{\ensuremath{\parallel}}
\newcommand{\orEr}{\ensuremath{|}}
\newcommand{\fail}{\ensuremath{\emptyset}}

\newcommand{\iriKind}{\ensuremath{\cM{IRI}}}

\newcommand{\Schema}{\ensuremath{\myfont{S}}}
\newcommand{\schema}{\Schema}

\newcommand{\AbstractSet}{\ensuremath{\myfont{A}}}
\newcommand{\schemaDef}{\ensuremath{\delta}}
\newcommand{\ShapeSet}{\ensuremath{\myfont{S}}}
\newcommand\IRISet{\myfont{I}}
\newcommand\BNodeSet{\myfont{B}}
\newcommand\LitSet{\mathit{Lit}}

\newcommand{\bnodeKind}{\ensuremath{\cM{BNode}}}
\newcommand{\datatype}[1]{\ensuremath{datatype(#1)}}
\newcommand{\bcond}[1]{\ensuremath{\Psi_{#1}}}
\newcommand{\predSpec}{\ensuremath{p}}

\newcommand{\openQs}[1]{\ensuremath{\lfloor#1\rfloor}}
\newcommand{\closeQs}[1]{\ensuremath{\lceil#1\rceil}}
\newcommand{\eachOfQs}[2]{#1\,,\,#2}
\newcommand{\oneOfQs}[2]{\ensuremath{#1\mid{}#2}}

\newcommand{\false}{\ensuremath{0}}
\newcommand{\iri}[1]{{\color{blue}\textit{:#1}}}
\newcommand{\iriq}[2]{{\color{red}#1}{\color{blue}\textit{:#2}}}
\newcommand{\literal}[1]{\ensuremath{\text{\texttt{#1}}}}
\newcommand{\extends}[2]{\ensuremath{extends\,@#1\,#2}}
\newcommand{\restricts}[2]{\ensuremath{restricts\,@#1\,#2}}
\newcommand{\trans}[2]{\ensuremath{\langle#2\rangle^{#1}}}
\newcommand{\transform}[2]{\trans{#1}{\iri{#2}}}
\newcommand{\embed}[3]{\ensuremath{\langle#2\rtimes#3\rangle^{#1}}}

\makeatletter
\newcount\my@repeat@count
\newcommand{\myrepeat}[2]{%
  \begingroup
  \my@repeat@count=\z@
  \@whilenum\my@repeat@count<#1\do{#2\advance\my@repeat@count\@ne}%
  \endgroup
}
\makeatother
\newcommand{\sep}[1]{\myrepeat{#1}{\space}}

\newcommand*{\SavedLstInline}{}
\LetLtxMacro\SavedLstInline\lstinline
\DeclareRobustCommand*{\lstinline}{%
  \ifmmode
    \let\SavedBGroup\bgroup
    \def\bgroup{%
      \let\bgroup\SavedBGroup
      \hbox\bgroup
    }%
  \fi
  \SavedLstInline
}

\lstdefinestyle{inlinecode}{basicstyle=\ttfamily\footnotesize\bfseries}
\renewcommand\c{\lstinline[style=inlinecode]}

\SetKwProg{Def}{def}{$\,$:}{}
\SetKwProg{Defn}{def}{~$=$}{}
\SetKw{defn}{def}
\newcommand{\DefInline}[2]{\defn #1 = #2}
\SetKwProg{DefnCustom}{\defn}{}{}
\SetKw{Let}{let}
\SetKwInput{KwIn}{Input}
\SetKwFor{ForEach}{foreach}{}{}
\SetKw{Or}{or}
\SetKw{And}{and}
\SetKwIF{If}{ElseIf}{Else}{if}{then}{else if}{else}{endif}
\SetKw{Match}{match}
\SetKw{MyIf}{if}
\SetKw{MyThen}{then}
\SetKw{MyElse}{else}
\SetKw{Case}{case}
\SetKwBlock{Let}{let}{in}
\SetKw{In}{in}
\SetKw{MapTo}{\ensuremath{\;\;\Rightarrow\;\;}}
\SetKwBlock{Block}{}{}
\newcommand{\assign}{\ensuremath{~\mathtt{:=}~}}
\newcommand{\algocomment}[1]{\text{//$\,$#1}}

\newcommand{\blockskip}{\smallskip}
\newcommand{\done}{\ensuremath{\mathit{done}}}

\def\algorithmsize{small}
\def\algorithmheadersize{\algorithmsize}
\renewcommand\AlCapFnt{\normalfont\bfseries\small}
\setlength{\textfloatsep}{1.0ex} 
\setlength{\floatsep}{1.0ex} 


\newcommand{\mycomment}[1]{\emph{// #1}}
\newcommand{\shapesGraph}{\ensuremath{S_g}}

\newcommand{\invPP}[1]{\ensuremath{~\hat{}#1}}
\newcommand{\seqPP}[2]{\ensuremath{#1\cdot{}#2}}
\newcommand{\altPP}[2]{\ensuremath{#1\lor{}#2}}
\newcommand{\notPP}[1]{\ensuremath{!#1}}
\newcommand{\starPP}[1]{\ensuremath{#1^{*}}}
\newcommand{\semPP}[2]{\ensuremath{|[#1|]^{#2}}}
\newcommand{\countProp}[6]{\ensuremath{\#_{#1,#6}^{#2,#3,#4,#5}}}
\newcommand{\countAll}[3]{\ensuremath{\#_{#1}^{#2,#3}}}
\newcommand{\conforms}[4]{#1,#2,#3\vDash#4}
\newcommand{\conformsNot}[4]{#1,#2,#3\nvDash#4}
\newcommand{\conformsTE}[4]{#1,#2,#3\Vdash#4}
\newcommand{\conformsQs}[4]{#1,#2,#3\vdash#4}

\newcommand{\slurpSE}[5]{#1,#2,#3\vDash#4\rightsquigarrow#5}
\newcommand{\slurpTE}[5]{#1,#2,#3\Vdash#4\rightsquigarrow#5}
\newcommand{\slurpQs}[5]{#1,#2,#3\vdash#4\rightsquigarrow#5}

\definecolor{forestgreen}{rgb}{34,139,34}
\definecolor{orangered}{rgb}{239,134,64}
\definecolor{darkblue}{rgb}{0.0,0.0,0.6}
\definecolor{gray}{rgb}{0.4,0.4,0.4}

\lstdefinestyle{SPARQL}{
 numberblanklines=false, 
 morekeywords={SELECT, FROM, WHERE, FILTER, GROUP BY, IN, AS, LIMIT,OFFSET,PREFIX,OPTIONAL,UNION,having,count,
 cex:,cdt:,dc:,dct:,dbo:,dbr:,doap:,ex:,foaf:,lemon:,org:,owl:,qb:,qp:,rdf:,rdfs:,sh:,sx:,schema:,skos:,skosxl:,void:,wr:,wt:,wf:,xsd:  
 },
 sensitive=false
}

\lstdefinestyle{Cypher}{
 numberblanklines=false, 
 morekeywords={MATCH, CREATE, RETURN, WITH, ORDER, BY, WHERE, FOREACH, DELETE, OPTIONAL, UNWIND, SKIP, LIMIT, SET, REMOVE, MERGE, CALL, UNION, USE, LOAD
 },
 sensitive=false
}

\lstdefinestyle{Turtle}{
  numberblanklines=false, 
  morekeywords={prefix, @prefix, 
 cex:,cdt:,dc:,dct:,dbo:,dbr:,doap:,ex:,foaf:,lemon:,org:,owl:,qb:,qp:,rdf:,rdfs:,sh:,sx:,schema:,skos:,skosxl:,void:,wr:,wt:,wf:,xsd:},
  alsoletter={:,@},
  comment=[l]{\#},
  morestring=[b]',
  morestring=[b]"
}

\lstdefinestyle{SHACL}{
    numberblanklines=false, 
    keywords={prefix, @prefix, 
   sh:and, sh:class, sh:closed, sh:constraintComponent, sh:conforms,
   sh:datatype, sh:disjoint, sh:equals, 
   sh:flags, sh:focusNode, 
   sh:hasValue,
   sh:ignoredProperties, sh:in, sh:js, sh:jsLibrary, sh:jsLibraryURL, sh:jsFunctionName, sh:JSConstraint,
   sh:languageIn, sh:lessThan, sh:lessThanOrEquals, 
   sh:maxCount, sh:maxExclusive, sh:maxInclusive, sh:maxLength, sh:minCount, sh:minExclusive, sh:minInclusive, sh:minLength, 
   sh:node, sh:nodeKind, sh:not, 
   sh:optional, sh:or, 
   sh:path, sh:pattern, sh:property, sh:qualifiedMaxCount, sh:qualifiedMinCount, sh:qualifiedValueShape,
   sh:qualifiedValueShape, sh:qualifiedValueShapesDisjoint, sh:sparql, 
   sh:resultSeverity, sh:resultPath, sh:resultMessage, 
   sh:sourceConstraintComponent, sh:sourceShape,
   sh:targetClass, sh:targetNode, sh:targetObjectsOf, sh:targetSubjectsOf,
   sh:uniqueLang, 
   sh:value, 
   sh:xone,
   sh:IRI, sh:Literal, sh:NonLiteral, sh:BlankNodeOrLiteral, sh:BlankNodeOrIRI, sh:IRIOrLiteral,
   sh:NodeShape, sh:PropertyShape, sh:ValidationReport, sh:ValidationResult, sh:Violation
   },
  morekeywords={
   cex:,cdt:,dc:,dct:,dbo:,dbr:,doap:,ex:,foaf:,lemon:,org:,owl:,qb:,qp:,rdf:,rdfs:,schema:,skos:,skosxl:,void:,wr:,wt:,wf:,xsd:,sh:,sx:
  }
  alsoletter={:,@},
  comment=[l]{\#},
  morestring=[b]',
  morestring=[b]",
  alsoletter={:}
}


\newcommand\JSONnumbervaluestyle{\color{blue}}
\newcommand\JSONstringvaluestyle{\color{red}}

\newif\ifcolonfoundonthisline

\makeatletter

\lstdefinestyle{json}
{
  showstringspaces    = false,
  keywords            = {false,true},
  alsoletter          = 0123456789.,
  morestring          = [s]{"}{"},
  stringstyle         = \ifcolonfoundonthisline\JSONstringvaluestyle\fi,
  MoreSelectCharTable =%
    \lst@DefSaveDef{`:}\colon@json{\processColon@json},
  basicstyle          = \ttfamily,
  keywordstyle        = \ttfamily\bfseries,
}

\newcommand\processColon@json{%
  \colon@json%
  \ifnum\lst@mode=\lst@Pmode%
    \global\colonfoundonthislinetrue%
  \fi
}

\lst@AddToHook{Output}{%
  \ifcolonfoundonthisline%
    \ifnum\lst@mode=\lst@Pmode%
      \def\lst@thestyle{\JSONnumbervaluestyle}%
    \fi
  \fi
 \lsthk@DetectKeywords%
}

\lst@AddToHook{EOL}%
  {\global\colonfoundonthislinefalse}

\makeatother
  
\lstdefinestyle{ShExC}{
  classoffset=0,
  keywords={start,
   VIRTUAL, CLOSED, EXTRA, IRI, BNode, Literal,
   NonLiteral, PATTERN, MININCLUSIVE, MINEXCLUSIVE, MAXINCLUSIVE, MAXEXCLUSIVE,
   TOTALDIGITS, FRACTIONDIGITS, LENGTH, MINLENGTH, MAXLENGTH, BASE, 
   AND, OR, NOT, true, false, end, let, prefix, 
   extends,
   restricts, 
   abstract},
  keywordstyle=\color{blue},
  classoffset=1,
  morekeywords={cex:,cdt:,
   dc:,dct:,dbo:,dbr:,doap:,
   ex:,foaf:,lemon:,org:,owl:,qb:,qp:,rdf:,rdfs:,
   sh:,sx:,schema:,skos:,skosxl:,void:,wr:,wt:,wf:,xsd:},
  keywordstyle=\color{green},
  classoffset=1,   
  morekeywords={class, export, boolean, throw, implements, import, this, js},
  keywordstyle=\color{darkgray}\bfseries,
  classoffset=0,
  identifierstyle=\color{black},
  morecomment=[s][\color{tealX}]{\%GenX\{}{\%\}},
  morecomment=[s][\color{tealJ}]{\%GenJ\{}{\%\}},
  morecomment=[s][\color{violet}]{\%js\{}{\%\}},
  morecomment=[s][\color{orange}]{\%sparql\{}{\%\}},
  sensitive=false,
  commentstyle=\color{purple}\ttfamily,
  stringstyle=\color{brown}\ttfamily,
  morestring=[b]',
  morestring=[b]",
  alsoletter={:},
  comment=[l]{\#}
  morecomment=[l]{//}
}

\lstdefinestyle{SHACLC}{
  keywords={closed,ignoredProperties,true,false,IRI,BlankNode,Literal,BlankNodeOrLiteral,IRIOrLiteral
  , prefix
   cex:,cdt:,dc:,dct:,dbo:,dbr:,doap:,ex:,foaf:,lemon:,org:,owl:,qb:,qp:,rdf:,rdfs:,sh:,sx:,schema:,skos:,skosxl:,void:,wr:,wt:,wf:,xsd:
},
 ndkeywordstyle=\color{darkgray}\bfseries,
  identifierstyle=\color{black},
  sensitive=false,
  commentstyle=\color{purple}\ttfamily,
  stringstyle=\color{brown}\ttfamily,
  morestring=[b]',
  morestring=[b]",
  alsoletter={:}
}

\lstdefinestyle{SQL}{numberblanklines=true, 
    morekeywords={CREATE,TABLE,ENUM,FOREIGN,KEY,ENUM,REFERENCES}}

\lstdefinestyle{HTML} {
    language=HTML,
    extendedchars=true, 
    breaklines=true,
    breakatwhitespace=true,
    emph={},
    emphstyle=\color{red},
    basicstyle=\ttfamily,
    columns=fullflexible,
    commentstyle=\color{gray}\upshape,
    morestring=[b]",
    morecomment=[s]{<?}{?>},
    morecomment=[s][\color{forestgreen}]{<!--}{-->},
    keywordstyle=\color{orangered},
    stringstyle=\ttfamily\color{orangered}\normalfont,
    tagstyle=\color{darkblue},
    morekeywords={attribute,xmlns,version,type,release},
    otherkeywords={attribute=, xmlns=},
}

\lstdefinestyle{XML} {
    language=XML,
    extendedchars=true, 
    breaklines=true,
    breakatwhitespace=true,
    emph={},
    emphstyle=\color{red},
    basicstyle=\ttfamily,
    columns=fullflexible,
    commentstyle=\color{gray}\upshape,
    morestring=[b]",
    morecomment=[s]{<?}{?>},
    morecomment=[s][\color{forestgreen}]{<!--}{-->},
    keywordstyle=\color{orangered},
    stringstyle=\ttfamily\color{orangered}\normalfont,
    tagstyle=\color{darkblue},
    morekeywords={attribute,xmlns,version,type,release},
    otherkeywords={attribute=, xmlns=},
}

\lstdefinestyle{RelaxNG} {
    extendedchars=true, 
    breaklines=true,
    breakatwhitespace=true,
    emph={},
    emphstyle=\color{red},
    basicstyle=\ttfamily,
    columns=fullflexible,
    commentstyle=\color{gray}\upshape,
    morestring=[b]",
    morecomment=[s]{<?}{?>},
    morecomment=[s][\color{forestgreen}]{<!--}{-->},
    keywordstyle=\color{darkblue},
    stringstyle=\ttfamily\color{orangered}\normalfont,
    keywords={attribute,element},
    otherkeywords={attribute=, xmlns=},
}

\lstdefinelanguage{JavaScript}{
  keywords={typeof, new, true, false, catch, function, return, null, catch, switch, var, if, in, while, do, else, case, break},
  keywordstyle=\color{blue}\bfseries,
  ndkeywords={class, export, boolean, throw, implements, import, this},
  ndkeywordstyle=\color{darkgray}\bfseries,
  identifierstyle=\color{black},
  sensitive=false,
  comment=[l]{//},
  morecomment=[s]{/*}{*/},
  commentstyle=\color{purple}\ttfamily,
  stringstyle=\color{red}\ttfamily,
  morestring=[b]',
  morestring=[b]"
}

\lstdefinelanguage{scala}{
  morekeywords={abstract,case,catch,class,def,%
    do,else,extends,false,final,finally,%
    for,if,implicit,import,match,mixin,%
    new,null,object,override,package,%
    private,protected,requires,return,sealed,%
    super,this,throw,trait,true,try,%
    type,val,var,while,with,yield},
  otherkeywords={=>,<-,<\%,<:,>:,\#,@},
  sensitive=true,
  morecomment=[l]{//},
  morecomment=[n]{/*}{*/},
  morestring=[b]",
  morestring=[b]',
  morestring=[b]"""
}

\definecolor{mygreen}{rgb}{0,0.6,0}
\definecolor{mygray}{rgb}{0.5,0.5,0.5}
\definecolor{mymauve}{rgb}{0.58,0,0.82}

\lstdefinestyle{inlinecode}{basicstyle=\ttfamily\footnotesize\bfseries}
\renewcommand\c{\lstinline[style=inlinecode]}

\lstset{
    basicstyle=\small\ttfamily,
    captionpos=b,
    frame=single,
    showstringspaces=false,
    escapeinside={\%*}{*)},
    style=ShExC
}


\newcommand{\hrefc}[3][blue]{\href{#2}{\color{#1}{#3}}}%

\newcommand{\elemento}[2]{\ensuremath{\hrefc[violet]{http://www.wikidata.org/entity/#2}{#1}}}
\newcommand{\propiedad}[2]{\ensuremath{\hrefc[darkblue]{http://www.wikidata.org/entity/#2}{#1}}}
\newcommand{\timBl}{\elemento{timBl}{Q80}}
\newcommand{\vintCerf}{\elemento{vintCerf}{Q92743}}
\newcommand{\newHaven}{\elemento{NewHaven}{Q49145}}
\newcommand{\bobDylan}{\elemento{bobDylan}{Q392}}
\newcommand{\London}{\elemento{London}{Q84}}
\newcommand{\CERN}{\elemento{CERN}{Q42944}}
\newcommand{\UK}{\elemento{UK}{Q145}}
\newcommand{\Spain}{\elemento{Spain}{Q29}}
\newcommand{\PA}{\elemento{PA}{Q329157}}
\newcommand{\birthDate}{\propiedad{birthDate}{P569}}
\newcommand{\birthPlace}{\propiedad{birthPlace}{P19}}
\newcommand{\country}{\propiedad{country}{P27}}
\newcommand{\employer}{\propiedad{employer}{P108}}
\newcommand{\awarded}{\propiedad{awarded}{P166}}
\newcommand{\pointTime}{\propiedad{pointTime}{P585}}
\newcommand{\start}{\propiedad{start}{P580}}
\newcommand{\pEnd}{\propiedad{end}{P582}}
\newcommand{\togetherWith}{\propiedad{togetherWith}{P1706}}
\newcommand{\Human}{\elemento{Human}{Q5}}
\newcommand{\instanceOf}{\propiedad{instanceOf}{P31}}

\newcommand{\indexn}[1]{\index{#1}#1}
\newcommand{\indexe}[1]{\index{#1}\emph{#1}}

\newcommand{\Ok}{\ensuremath{\c|Ok|}}
\newcommand{\Failed}{\ensuremath{\c|Failed|}}
\newcommand{\WaitingFor}[3]{\ensuremath{\c|WaitingFor|(#1,#2,#3)}}
\newcommand{\Pending}{\ensuremath{\c|Pending|}}
\newcommand{\PendingLs}{\ensuremath{\c|Pending|(ls)}}
\newcommand{\Undefined}{\ensuremath{\c|Undefined|}}

\newcommand{\Validate}{\ensuremath{\c|Validate|}}
\newcommand{\Checked}[2]{\ensuremath{\c|Checked|(#1,#2)}}
\newcommand{\WaitFor}[1]{\ensuremath{\c|WaitFor|(#1)}}
\newcommand{\status}[2]{\ensuremath{#1(#2)}}
\newcommand{\msgSent}[3]{\ensuremath{#1,#2\rightsquigarrow{}#3}}
\newcommand{\checkLocal}[2]{\ensuremath{\c|checkLocal|(#1,#2)}}
\newcommand{\checkLocalOpen}[2]{\ensuremath{\c|checkLocalOpen|(#1,#2)}}

\newcommand{\checkNeighs}[3]{\ensuremath{\c|checkNeighs|(#1,#2,#3)}}
\newcommand{\fracEmpty}[2]{\genfrac{}{}{0pt}{0}{#1}{#2}}
\newcommand{\vProg}{\c|vProg|}
\newcommand{\tripleConstraints}[1]{\c|tripleConstraints(#1)|}
\newcommand{\rbe}[1]{\ensuremath{\c|rbe|(#1)}}
\newcommand{\combine}[2]{\ensuremath{\c|combine|(#1,#2)}}

%% file: 10_introduction.tex
\section{Introduction} \label{sec:intro}

\index{Google} 
The concept of Knowledge Graphs was popularized by Google in 2012~\cite{GoogleKG} 
 as a tool that collects information about real world entities and makes  relationships between them 
 with the goal of improving search results, 
 understand relationships better
 and make unexpected discoveries.
Since them, there has been a tremendous interest and adoption about Knowledge Graphs, with open, general purpose ones 
 as well as closed, proprietary ones like those employed by some big companies.
 \index{DBpedia} \index{YAGO} \index{Wikidata}
In the former case we can mention DBpedia~\cite{auer2007dbpedia}, YAGO~\cite{suchanek2007yago} or Wikidata~\cite{VrandecicK14}.
In the latter, some example companies that have announced their use are
\index{Airbnb} Airbnb~\cite{AirBnBKG}, 
\index{Amazon} Amazon~\cite{AmazonKG}, 
\index{eBay} eBay~\cite{eBayKG}, 
\index{Facebook} Facebook~\cite{NoyGJNPT19}, 
\index{IBM} IBM~\cite{IBMKG}, 
\index{LinkedIn} LinkedIn~\cite{LinkedInKG}, 
\index{Microsoft} Microsoft~\cite{BingKG}, etc.


There are different models associated with Knowledge Graphs like
 RDF-based graphs, property graphs and wikibase graphs:
 
 \begin{itemize}
\index{RDF-based graphs} 
\item RDF-based graphs is one of the most well-known data models 
given that RDF was proposed as a W3C recommendation already in 1999~\cite{RDF10_99} 
and a large ecosystem of tools have been created around it. 
 An important aspect of RDF is the use of URIs, which facilitates interoperability and was the basis semantic web and linked data.
 Around RDF, a whole ecosystem of technologies have appeared, like the SPARQL query language and protocol~\cite{sparql11}, which enables the creation of public endpoints.
 
\index{New4j} \index{Property graphs} 
\item Graph databases like Neo4j~\cite{Miller13} 
 have also been employed to 
 represent knowledge graphs. 
 They have a data model which allows to annotate 
 both nodes and edges with pairs of property-values which has become to be known as property graphs~\cite{Rodriguez2010}.
 
\index{Wikidata} \index{Wikipedia} 
\item Wikidata started in 2012 as a support project for Wikipedia but has been  
 evolving and acquiring more and more importance as a 
 hub of public knwowledge.
 The data model emplyed by Wikidata combined several aspects from RDF following linked data principles and from property graphs, allowing the annotation of statements by property-values using qualifiers and references. 
 The software suite that implements Wikidata is known as Wikibase and can be used to represent other knowledge graphs with the same data model, we will refer to these kind of knowledge graphs as wikibase graphs.
 Wikidata also offers an RDF serialization format which can be accessed through a public SPARQL endpoint.
\end{itemize} 
 
One of the key factors of knowledge graph models is their flexibility which enables easy addition of content. 
This flexibility also comes with a price for the applications and users that want to consume the data, 
which are required to use defensive programming techniques to handle lack of some mandatory properties, errors in values, duplicates, etc.
It is also difficult for the producers who want to add data. 
Although they usually have some schema (explicit or implicit) about that represents the structure of the data, 
it is also difficult to document that the added data conforms to that schema.
 
\index{Entity schemas} \index{ShEx} \index{Shape Expressions} \index{RDF serialization} 
In the case of RDF graphs, Shape Expressions (ShEx) were developed in 2014 to describe and validate the topology of RDF graphs~\cite{Prudhommeaux2014}. 
Afterwards, ShEx was adopted in 2019 by Wikidata to describe the RDF serialization of Wikidata content in a new namespace called entity schemas\footnote{\url{https://www.wikidata.org/wiki/Wikidata:Schemas}}.

\begin{figure}
\centering
\begin{tikzpicture}
\begin{axis}[
    title={Size of Wikidata dumps},
    xlabel={Year},
    ylabel={Size (Gigabytes)},
    xmin=14, xmax=21,
    ymin=0, ymax=100,
    xtick={14,15,16,17,18,19,20,21},
    ytick={0,20,40,60,80,100},
    legend pos=north west,
    ymajorgrids=true,
    grid style=dashed,
]

\addplot[
    color=blue,
    mark=square,
    ]
    coordinates {
    (14,3.3)(15,4.8)(16,9.9)(17,19.2)(18,24.1)(19,36.3)(20,58.4)(21,70.5)
    };
    \legend{Size JSON dumps (Gb)}
\end{axis}
\end{tikzpicture}
\caption{Size of wikidata Json Dumps between 2014-2021. Source: \href{https://archive.org/details/wikimediadownloads?and\%5B\%5D=\%22Wikidata\%20entity\%20dumps\%22}{http://archive.org}} \label{fig:sizeEvolutionWikidata}
\end{figure}
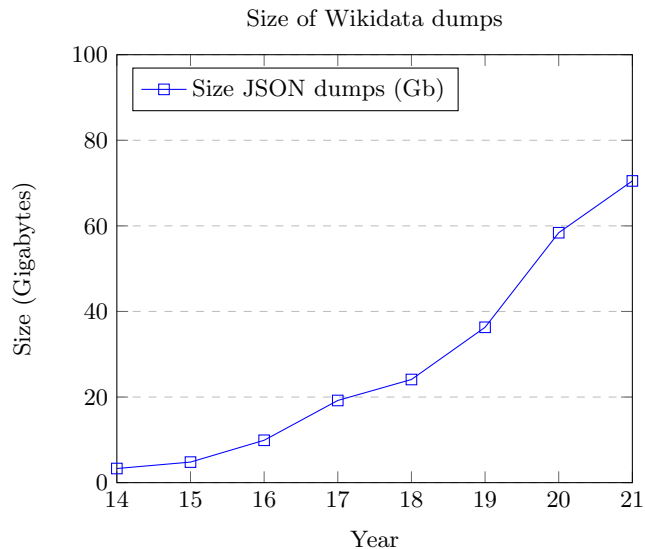

The success of Knowledge Graphs has implied that the size of their contents also increases dramatically. 
 As an example, the size of compressed Wikidata dumps has been almost doubling every year, from 3.3Gb (31,3Gb uncompressed) in 2014 to 70.5Gb (1.256Gb uncompressed) in 2021 (see figure ~\ref{fig:sizeEvolutionWikidata}).
 A consequence of these huge sizes is that it is not possible to easily process all the amount of available data by conventional tools, preventing the consumers to analyze and process the content and threatens these technologies to be victims of their own success.
 
 \index{Scholia}
 An example of this situation happens in Scholia~\cite{Nielsen2017} a web application that leverages on Wikidata to represent information about scholars and their works. The application also contains nice visualizations and comparisons which are based on queries over Wikidata endpoint. Although the project provides a lot of interesting information, the more complex visualizations are not possible to obtain because the huge amounts of data generate timeouts. 
 
In order to address these issues, a possible approach is to create subsets of the Knowledge Graphs for some domains. 
These subsets can capture snapshots of the content at some specific moment and be used to improve the performance of the applications that consume that data, facilitating research work over Knowledge Graphs contents.

In this paper we review different approaches to generate knowledge graphs subsets. Given that the first step to create a subset is to describe
the intended content, ShEx seems a natural choice to be used for those descriptions. In this way, some approaches are based on ShEx schemas which describe the intended content of the subsets. 
\index{PShEx} \index{WShEx}
To that end, we define two extensions of ShEx: PShEx to describe property graphs and WShEx to describe wikibase graphs.

We define the following approaches to generate subsets for Wikibase graphs which could also be applied to RDF graphs and property graphs:

\begin{itemize}
\item \emph{Entity-matching} defines a subset by identifying some target entities. The subset contains information related with those entities and their neighborhood.

\item\emph{Simple-matching} defines a subset by a set of matching patterns, for example, the triples that have a given property, that satisfy some condition, etc. 

\item\emph{ShEx-based matching} uses ShEx shapes without taking into account shape references to check which nodes conform to them. This approach only requires to take into account the neighborhood of a node and can be used to sequentially process the dumps without requiring graph traversal. 

\index{ShEx+Slurp}
\item\emph{ShEx+Slurp} consists on validating the graph contents using ShEx and collect the nodes and triples that are being visited during the validation. This approach can refine the obtained subsets
but requires graph traversal, which can be difficult when sequentially processing the dumps. If it is used against an endpoint, it can exceed the limit of allowed requests by client.
 
\index{ShEx+Pregel} 
\item\emph{ShEx+Pregel} proposes to validate the graph using an adaptation of the Pregel algorithm~\cite{Malewicz2010} for ShEx validation. 
This approach can process and validate big knowledge graphs. 
 It has the advantage of scalability and in principle, 
 it can handle graph traversal, but it also consumes a large amount of resources.
\end{itemize}

The main contributions of this paper are:
\begin{itemize}
\item We created a formal model for Wikibase graphs which can be compared with the formal model of RDF-graphs and property graphs.
\item We created two extensions of ShEx: PShEx for property graphs and WShEx for wikibase graphs.
\item We identify and formally describe five approaches to generate knowledge graphs subsets. Some of them, like \emph{Simple matching} and \emph{ShEx+Slurp} had already been implemented but were not formally described.
\item We describe and implement an algorithm for large scale validation of knowledge graphs based on Pregel.
\end{itemize}

The structure of the paper is as follows:
Section~\ref{sec:preliminaries} presents some preliminary definitions about sets and graphs.
Section~\ref{sec:KnowledgeGraphs} introduces knowledge graphs and presents 3 main types of knowledge graphs: RDF-based, Property graphs and Wikibase graphs.
Section~\ref{sec:DescribingKGs} presents techniques to describe knowledge graphs: ShEx for RDF graphs, PShEx for property graphs and WShEx for wikibase graphs. For each of them, we define the abstract syntax and the semantics using inference rules.
Section~\ref{sec:KGSubSets} presents the problem of creating subsets of knowledge graphs and describes several approaches to create subsets of wikibase graphs. 
Finally, section~\ref{sec:RelatedWork} reviews the related work and section~\ref{sec:Conclusions} presents some conclusions.
\index{Tim Berners-Lee}
Along the paper we use a running example based on information about Tim Berners-Lee whose information was obtained from Wikidata (entity \href{http://www.wikidata.org/entity/Q80}{Q80}).

%% file: 15_preliminaries.tex
\section{Preliminaries} \label{sec:preliminaries}

In this section, we provide some basic definitions that we will use in the rest of the paper.

\index{Set} \index{Cartesian product}
\paragraph{Sets}. 
The finite set with elements $a_1, \ldots, a_n$ is written $\{a_1, \ldots, a_n\}$,
$\emptyset$ represents the empty set,
$S_1\cup{}S_2$ is the union of sets $S_1$ and $S_2$,
$S_1\cap{}S_2$ the intersection and 
$S_1\times{}S_2$ the Cartesian product.
$\FinSet{S}$ represents the set of all finite subsets of $S$.
A tuple $\langle{}A_1,\dots{}A_n\rangle$ is the cartesian product $A_1\times\dots\times{}A_n$.

Given a set $S$, its set of partitions is defined as 
 $\partition{s}=\{(s_1,s_2) \mid s_1\cup{}s_2=s\wedge{}s_1\cap{}s_2=\emptyset\}$.

\index{Graph} 
\begin{definition}[Graph] \label{def:graph}
A \emph{graph} is a tuple $G = \langle\VertSet,\EdgeSet\rangle$, where $\VertSet$ is a set of nodes, and 
 $\EdgeSet \subseteq \VertSet\times\EdgeSet\times\VertSet$ 
 is a set of edges.
\end{definition}

A \indexn{multigraph} is a graph where it is possible to have more than one edge between the same two nodes.

\index{Directed edge-labelled graph}
\begin{definition}[Directed edge-labelled graph]\label{def:delg}
A \emph{directed edge-labelled graph} is a tuple 
$\Graph = \langle\VertSet,\EdgeSet,\PropSet\rangle$, where $\VertSet$ is a set of nodes, $\PropSet$ is a set of labels also called predicates or properties, and $\EdgeSet \subseteq \VertSet \times \LabelSet \times \VertSet$ is a set of edges. 
\index{triple}
\index{subject}
\index{predicate}
\index{object}
Each element $(x,p,y)\in\EdgeSet$ is called a triple, where $x$ is the subject, $p$ is the predicate or property and $y$ is the object.
\end{definition}

\index{triple-based graphs}
\begin{definition}[Triple-based graphs]
A \emph{triple-based graphs} is a directed edge-labelled graph 
$\Graph = \langle\SubjectSet,\PropSet,\ObjectSet,\rho\rangle$ where 
$\SubjectSet$ is a set of subjects, 
$\PropSet$ is a set of predicates or properties and 
$\ObjectSet$ is a set of objects or values, and $\rho\subseteq\SubjectSet\times\PropSet\times\ObjectSet$. 
Those sets don't need to be disjoint, and 
usually $\PropSet\subseteq\SubjectSet\subseteq\ObjectSet$.
\end{definition}

\index{hypergraph}
\begin{definition}[Hypergraph] \label{Hypergraph}
A \emph{hypergraph} is a tuple $G=\langle\VertSet,\EdgeSet\rangle$ where $\VertSet$ is a set of nodes and $\EdgeSet\subseteq\FinSet{\VertSet}$ is a set of edges. Notice that $\EdgeSet$ is a family of subsets of vertices.
\end{definition}

\newcommand{\cupplus}{\uplus}

\begin{definition}[Bag~\cite{Slawek2015}] \label{Bag}
\index{bag} \index{multiset} \index{bag language}
Given a set of symbols $\Delta$, a \emph{bag} $\bagw$ (also called \emph{multiset}) can be seen as a set whose elements can be repeated and it is defined as a function $\bagw:\Delta\mapsto\Nat$ that maps a symbol to the number of its occurrences. The set of all bags over $\Delta$ is denoted as $Bag[\Delta]$.
 The empty bag $\emptyBag$ has 0 occurrences for every symbol $a\in\Delta$. 
 A bag is usually represented as $\bag{a\dots{}}$ with elements that can be repeated. 
 The union of two bags $\bagw_1$ and $\bagw_2$ is defined as $\bagw_1\cupplus\bagw_2(a)=\bagw_1(a)+ \bagw_2(a)$.
 A set of bags is a \emph{bag language}. 
 The bag union of two languages $L_1$ and $L_2$ is defined as 
 $L_1\cupplus{}L_2=\{\bagw_1\cupplus{}\bagw_2\mid\bagw_1\in{}L_1, \bagw_2\in{}L_2\}$.
\end{definition}

\index{Regular bag expression}
\begin{definition}[Regular bag expression~\cite{Slawek2015}]
A \emph{regular bag expression} over a set of symbols $\Delta$ is defined by the following grammar:
 $E::=\emptyEr\big\vert{}a\big\vert{}E\mid{}E\big\vert{}E;E\big\vert{}E*$, where $a\in\Delta$.  

 A regular bag expression $e$ defines a bag language $\bagSem{e}$ whose semantics is:
 $\bagSem{\emptyEr}=\{\epsilon\}, 
  \bagSem{a}=\{\bag{a}\}, 
  \bagSem{e_1\mid{}e_2}=\bagSem{e_1}\cup\bagSem{e_2}, 
  \bagSem{e_1;e_2}=\bagSem{e_1}\cupplus\bagSem{e_2},
  \bagSem{e*}=\bigcup_{i\geq0}\bagSem{e}^i$.
A bag $b$ matches a regular bag expression $e$, denoted as $\matchRbe{b}{e}$ if $b\in{}\bagSem{e}$.   

\end{definition}





\index{Shape assignment}
\begin{definition}[Shape assignment] \label{Shape assignment}
Given a graph $\Graph$ with vertex set $\VertSet$ and a finite set of labels $\LabelSet$, 
a \emph{shape assignment} over $\Graph$ and $\LabelSet$ is a subset of $\VertSet\times\LabelSet$.
We use $\typing$ to denote shape assignments, and we write $\hasType{n}{\lbl}$ instead of $(n,\lbl)$ for elements of shape assignments.
\index{shape maps} \index{typings}
Note that shape assignments correspond to shape maps in~\cite{ShExSpec} and typings in~\cite{Boneva17,Slawek2015}.
\end{definition}

%% file: 20_KnowledgeGraphs.tex
\section{Knowledge graphs models} \label{sec:KnowledgeGraphs}

\index{Knowledge Graph}
Although the term \emph{knowledge graphs} was already in use in the 1970s~\cite{Schneider72}, 
the current notion of knowledge graphs was popularized by 
\indexn{Google} in 2012~\cite{GoogleKG}. 
We adopt an informal definition of knowledge graphs which has been inspired by Hogan et al~\cite{Hogan2021}:

\begin{definition}[Knowledge graph]
A \emph{Knowledge graph} is graph of data intended to represent knowledge of some real world domain, 
whose nodes represent entities of interest and 
whose edges represent relations between these entities.
\end{definition}

The previous definition is deliberately open. 
The main feature of a knowledge graph is that it is intended to represent 
 information about entities of some real world domain using 
 a graph-based data structure. 

Knowledge graphs are usually classified by:

\begin{itemize}
	\item Licence/proprietor: 
	There are public and open knowledge graphs like 
	\indexn{Yago}~\cite{suchanek2007yago}, 
	\indexn{DBpedia}~\cite{LehmannIJJKMHMK15} or 
	\indexn{Wikidata}~\cite{VrandecicK14} as well as 
	enterprise-based and proprietary knowledge graphs~\cite{NoyGJNPT19} like 
	\indexn{Google}, 
	\indexn{Amazon}, etc. 
	
	\item Scope: there are general-purpose knowledge graphs which contain information about almost 
	all domains like \indexn{Wikidata} as well as 
	domain specific knowledge graphs which contain information from some specific domains like 
	\indexn{healthcare}, 
	\indexn{education}, 
	\indexn{chemistry}, 
	\indexn{biology}, 
	\indexn{cybersecurity}, etc.~\cite{AbuSalih2021}
\end{itemize}

Knowledge graphs can be represented using multiple technologies and in fact, the information about how 
Google's knowledge graph is implemented is not public. 
Nevertheless, in this paper, we will focus on three main technologies:
\begin{itemize}
    \index{RDF} 
	\item \textbf{RDF based knowledge graphs} represent information using 
	  directed graphs whose edges are labels. 

    \index{Property graphs}
	\item \textbf{Property graphs} allow property–value pairs and a label to be associated with nodes and edges. 
	  Property graphs have been implemented by several popular 
	  \indexn{graph databases} like \indexn{Neo4j}~\cite{AnglesABHRV17}.
	
	\index{Attributed graphs}
	\item \textbf{Attributed graphs} allow property-value pairs associated with edges to add meta-data about the relationship represented by the edge and the values of those properties can themselves be nodes in the graph.
    The main example in this category is \indexn{Wikidata} where property-value pairs encode 
    qualifiers and references.
\end{itemize}

\subsection{RDF based knowledge graphs}

\index{RDF}
Resource Description Framework (RDF)~\cite{rdf11}, 
 is a \indexn{W3C recommendation} which is based on directed edge-labelled graphs. 

The \indexn{RDF data model} defines different types of nodes, 
including 
\indexe{Internationalized Resource Identifiers} (\indexe{IRIs})~\cite{rfc3987} which can be used to 
 globally identify entities on the Web; 
\indexn{literals}, which allow for representing strings (with or without \indexn{language tags}) 
 and values from other \indexn{datatypes} (\indexn{integers}, \indexn{decimals}, \indexn{dates}, etc.); 
and \indexe{blank nodes}.
Blank nodes can be considered as existential variables that denote 
 the existence of some resource for which an IRI or literal is not known or provided. 
They are locally scoped to the file or \indexn{RDF store}, 
 and are not persistent or have portable identifiers~\cite{HoganAMP14}. 

\begin{definition}[RDF Graph]
Given a set of \indexn{IRIs} $\IRISet$, 
 a set of \indexn{blank nodes} $\BNodeSet$ and 
 a set of \indexn{literals} $\LitSet$, 
an \indexe{RDF graph} is a \indexn{triple based graph}
$\Graph = \langle\SubjectSet,\PropSet,\ObjectSet,\StmtSet\rangle$ 
where 
$\SubjectSet=\IRISet\cup\BNodeSet$,
$\PropSet=\IRISet$, 
$\ObjectSet=\IRISet\cup\BNodeSet\cup\LitSet$ and 
$\StmtSet\subseteq\SubjectSet\times\PropSet\times\ObjectSet$
\end{definition}

There are several syntaxes for RDF graphs like \indexn{Turtle}, \indexn{N3}, \indexn{RDF/XML}, etc. 
 In this document, we will use Turtle.

\begin{example} \label{example:RDFData}
	
As a running example, 
 we will represent information about \indexn{Tim Berners-Lee} 
 declaring that he was born in \indexn{London}, 
 on 1955, and was employed by \indexn{CERN} and London's country is UK.  
 That information can be encoded in Turtle as:

\begin{lstlisting}[style=Turtle]
prefix :      <http://example.org/>
prefix xsd:   <http://www.w3.org/2001/XMLSchema#>

:timbl   :birthPlace     :London ;
         :birthDate      "1955-08-06"^^xsd:date ;
         :employer       :CERN .
:London  :country        :UK .
\end{lstlisting}


Figure~\ref{figure:RDFExample} shows a possible visualization of that RDF graph using \indexn{RDFShape}, 
a tool developed by the authors of this paper which allows to play with 
RDF graphs\footnote{It is possible to interactively play with the example following this permalink:~\url{https://rdfshape.weso.es/link/16344135752}}~\cite{RDFShape18}:

\begin{figure}
\centering
\includegraphics[width=0.5\textwidth]{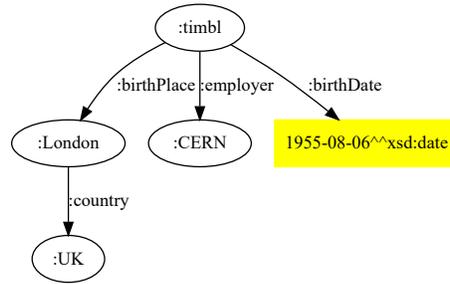}
\caption{Example graph representation of RDF data} \label{figure:RDFExample}
\end{figure}

\end{example}

The neighbors of a node $n\in\VertSet$ in an RDF graph $\Graph$ are defined as 
 $neighs(n,\Graph)=\{(n,p,y)\mid (n,p,y)\in\Graph \}$. 
 
RDF can be considered the basic element of the semantic web technology stack, 
 forming a simple knowledge representation language on top of which several technologies 
 have been developed like \indexn{SPARQL} for querying RDF data as well as 
 \indexn{RDFS} and \indexn{OWL} to describe vocabularies and ontologies.

\paragraph{RDF reification and RDF-*}

An important aspect of \indexn{RDF} as a knowledge representation formalism is to be able to represent 
 information about RDF triples themselves, which is called \indexn{reification}. 
In this section we will present some of the possible approaches for reification 
 using a simple example to help understand the approach used by \indexn{Wikibase} 
  to serialize its data model to RDF~\cite{HHK2015}.
We also present the \indexn{RDF-*} approach which has become popular in the 
 RDF-ecosystem with its support by several RDF stores like \indexn{GraphDB}~\footnote{\url{https://www.ontotext.com/knowledgehub/fundamentals/what-is-rdf-star/}}.

\begin{example}
As an example, we may want to qualify the statement that Tim Berners-Lee was employed by CERN, declaring that 
he was employed at two different points in time: 
in 1980 and between 1984 and 1994. 



The following approaches have been proposed for RDF reification:

\begin{itemize}
\item \indexn{Standard RDF reification} was introduced in \indexn{RDF 1.0}~\cite{rdf10}. 
 It consists of using the predicates \c|rdf:subject|, \c|rdf:predicate| and \c|rdf:object| 
  as well as the class \c|rdf:Statement| to explicitly declare statements.

\begin{lstlisting}[style=Turtle]
_:s1 rdf:type        rdf:Statement ;
     rdf:subject     :timbl ;
     rdf:predicate   :employer ;
     rdf:object      :CERN ;
     :start          "1980"^^xsd:gYear ;
     :end            "1980"^^xsd:gYear .
_:s2 rdf:type        rdf:Statement ;
     rdf:subject     :timbl ;
     rdf:predicate   :employer ;
     rdf:object      :CERN ;
     :start          "1984"^^xsd:gYear ;
     :end            "1994"^^xsd:gYear .
\end{lstlisting}

\item Create a statement that models the $n$-ary relation~\cite{EGKMV2014}.
 For example, we can create two nodes \lstinline|:s1| and \lstinline|:s2| to represent the
  the 2 employments of Tim-Berners-Lee at CERN.

\begin{lstlisting}[style=Turtle]
:timbl :employer :s1, :s2.
:s1 :employerV :CERN;
    :start     "1980"^^xsd:gYear ;
    :end       "1980"^^xsd:gYear .
:s2 :employerV :CERN;
    :start     "1984"^^xsd:gYear ;
    :end       "1994"^^xsd:gYear .
\end{lstlisting}

\item Create \indexe{singleton properties} for each statement and link those properties with 
 a specific predicate to the real property~\cite{Nguyen2014}.

\begin{lstlisting}[style=Turtle]
:timbl :employer1 :CERN ;
       :employer2 :CERN .
       
:employer1 :singletonPropertyOf :employer ;
 :start "1980"^^xsd:date ;
 :end   "1980"^^xsd:date .
 
:employer2 :singletonPropertyOf :employer ;
 :start "1984"^^xsd:date ;
 :end   "1994"^^xsd:date .
\end{lstlisting}

\item RDF1.1~\cite{rdf11} included the concept of \indexn{named graphs}, which can be used to associate each triple with a different graph.

\begin{lstlisting}[style=Turtle]
:g1 :timbl    :employer :CERN .
:g1 :employed :start    "1980"^^xsd:date  .
:g1 :employed :end      "1980"^^xsd:date  .
:g2 :timbl    :employer :CERN .
:g2 :employed :start    "1984"^^xsd:date  .
:g2 :employed :end      "1994"^^xsd:date  .
\end{lstlisting}

\item \indexn{RDF-*}~\cite{rdfstar} has been recently introduced as an extension of RDF that includes 
 RDF graphs as either the subjects or objects of a statement. 

\begin{lstlisting}[style=Turtle]
<<:timbl :employer :CERN>> :start "1980"^^xsd:gYear ;
                           :end   "1980"^^xsd:gYear .
<<:timbl :employer :CERN>> :start "1984"^^xsd:gYear ;
                           :end   "1994"^^xsd:gYear .
\end{lstlisting}

\item Wikidata's RDF serialization follows a hybrid approach using 
 a direct link to capture the preferred value and 
 singleton nodes that represent the statements capturing the $n$-ary relationship~\cite{EGKMV2014}. 
 It also follows a convention that employs the same local name of the property preceded by different 
  namespaces: \lstinline|wdt:| for the direct link, \lstinline|p| for the link between the node and 
  the singleton statements, \lstinline|ps:| for the link between the singleton statements and the values,
  and \lstinline|pq:| for the link between the singleton statements and the qualified values. 
  The previous example using Wikidata RDF serialization could 
  be~\footnote{We omit the representation of values and use English names instead of numbers for clarity}:

\begin{lstlisting}[style=Turtle]
:timbl wdt:employer :CERN .
:timbl p:employer :s1 .
:timbl p:employer :s2 .
:s1    ps:employer :CERN ;
       pq:start "1980"^^xsd:gYear ;
       pq:end   "1980"^^xsd:gYear .
:s2    ps:employer :CERN ;
       pq:start "1984"^^xsd:gYear ;
       pq:end   "1994"^^xsd:gYear .
\end{lstlisting}
\end{itemize}
\end{example}

\subsection{Property graphs} \label{sec:PropertyGraphs}

\index{Property graphs}
Property graphs have become popular thanks to several commercial graph databases like 
\indexn{Neo4j}~\footnote{\url{https://neo4j.com/}}, 
\indexn{JanusGraph}~\footnote{\url{https://janusgraph.org/}} or 
\indexn{Sparksee}~\footnote{\url{https://www.sparsity-technologies.com/\#sparksee}}. 
A property graph has \indexn{unique identifiers} for each node/edge and allows to add \indexn{property-value annotations} to each node/edge in the arc as well as \indexn{type annotations}.

The following definition of a property graph follows~\cite{seifer2021progs}.

\begin{definition}[Property graph]
Given a set of types $\PLabelSet$,
a set of properties $\PropSet$, 
and a set of values $\ValueSet$,
a \emph{property graph} $\Graph$ is a tuple 
 $\langle\NodeSet,\EdgeSet,\rho,\lambda_n,\lambda_e,\sigma\rangle$ where 
 $\NodeSet\cap\EdgeSet=\emptyset$,
 $\rho:\EdgeSet\mapsto\NodeSet\times\NodeSet$ is a total function,
 $\lambda_n:\NodeSet\mapsto\FinSet{\PLabelSet}$, 
 $\lambda_e:\EdgeSet\mapsto\PLabelSet$, and
 $\sigma:\NodeSet\cup\EdgeSet\times\PropSet\mapsto\FinSet{\ValueSet}$. 
\end{definition}

A property graph is formed by a set of node identifiers $\NodeSet$
 and a set of edges $\EdgeSet$ where 
 $\rho$ associates a pair of nodes $(n_1,n_2)$ to every $e\in\EdgeSet$ where $n_1$ is the subject and $n_2$ is the object,
 $\lambda_n$ associates a set of types for node identifiers (notice that property graphs allow nodes to have more than one type),
 $\lambda_e$ associates a types for each edge identifier,
 and 
 $\sigma$ associates a set of values to pairs $(i,p)$ such that $i\in\NodeSet\cup\EdgeSet$ is a node or edge and $p\in\PropSet$ is a property.
 
\begin{example}\label{example:PropertyGraphs}

As an example, we will represent information about Tim Berners-Lee in a property graph 
encoding his birth place with a relation to a node that represents London, and his birth date with a value for that property in the same node. 
We can also represent that its employer has been CERN in two times, 
 one in 1980, and another between 1984 and 1994. 

\begin{flalign*}
\PLabelSet &= \{\text{Human},\text{City},\text{Metropolis},\text{Country},\text{Organization},\text{birthPlace},\text{country},\text{employer} \} &\\
\PropSet &= \{\text{label},\text{birthDate},\text{start},\text{end} \} &\\
\ValueSet &= \{\text{Tim Berners-Lee},\text{1955},\text{1980},\text{1984},\text{1994},\text{London},\text{UK} \} &\\
\NodeSet &= \{n_1,n_2,n_3,n_4\}\qquad \EdgeSet=\{r_1,r_2,r_3,r_4\} &\\
\rho &= r_1\mapsto(n_1,n_2), r_2\mapsto(n_2,n_3), r_3\mapsto(n_1,n_4), r_4\mapsto(n_1,n_4) &\\
\lambda_n &=n_1\mapsto{}\{\text{Human}\},\,n_2\mapsto{}\{\text{City,Metropolis}\},\,n_3\mapsto{}\{\text{Country}\},\,n_4\mapsto{}\{\text{Organization}\}&\\
\lambda_e &=r_1\mapsto{}\text{birthPlace},r_2\mapsto{}\text{country},r_3\mapsto{}\text{employer},r_4\mapsto{}\text{employer} &\\
\sigma &=(n_1,label)\mapsto{}\text{Tim Berners-Lee},\,(n_1,birthDate)\mapsto{}\text{1955} &\\
 & (n_2,\text{label})\mapsto{}\text{London}\},\,(n_3,\text{label})\mapsto{UK},\,(n_4,\text{label})\mapsto{}\text{CERN} &\\
 & (r_3,\text{start})\mapsto{}\text{1980},\,(r_3,\text{end})\mapsto{}\text{1980},\,(r_4,\text{start})\mapsto{}\text{1984},\,(r_4,\text{end})\mapsto{}\text{1994}
\end{flalign*}
\end{example}

\begin{figure}[h!]
\centering
\includegraphics[width=0.9\textwidth]{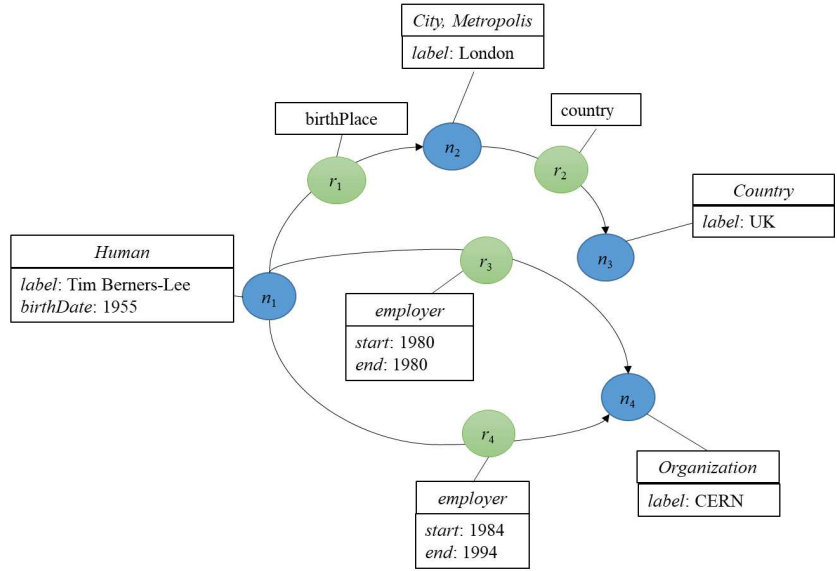}
\caption{Example graph visualization of a property graph} \label{figure:PropertyGraphExample}
\end{figure}

Figure~\ref{figure:PropertyGraphExample} presents a possible visualization of a property graph.

\indexn{Cypher} is a property graph query language that was initially developed for Neo4j~\cite{Francis2018}. 
Figure~\ref{example:cypher} presents an example Cypher script that can can generate the property graph represented in figure~\ref{figure:PropertyGraphExample}.

\begin{figure}[h!]
\begin{lstlisting}[style=Cypher]
CREATE (n1:Human {label:'Tim Berners-Lee', birthDate:1955})
CREATE (n2:City:Metropolis {label:'London'})
CREATE (n3:Country {label:'UK'})
CREATE (n4:Organization {label:'CERN'})
CREATE
  (n1)-[:birthPlace]->(n2),
  (n2)-[:country]->(n3),
  (n1)-[:employer {start:[1980], end: [1980]}]->(n4),
  (n1)-[:employer {start:[1984], end:[1994]}]->(n4),
\end{lstlisting}
\caption{Cypher code to generate a sample property graph} \label{example:cypher}
\end{figure}
Notice that it is possible to have more than one edge between nodes in property graphs, so they can be considered multigraphs.

\subsection{Wikibase graphs}

\indexn{Wikidata}\footnote{\url{http://wikidata.org/}} started in 2012 
 to support \indexn{Wikipedia}~\cite{VrandecicK14}. 
It has become one of the biggest human knowledge bases, 
 maintained both by humans collaboratively as by bots, 
 which update the contents from external services or databases. 
Several organizations are donating their data to Wikidata and collaborate in its maintenance providing resources. 
\index{Freebase} \index{Google}
A remarkable case is Google, 
 which migrated its previous knowledge graph Freebase to Wikidata in 2017~\cite{Tanon2016}.

\index{Wikidata} \index{Wikipedia}
Apart of Wikipedia, 
 Wikidata has been reported to be used by external applications like 
 \index{Apple} Apple's \index{Siri} Siri~\footnote{\url{https://lists.wikimedia.org/pipermail/wikidata/2017-July/010919.html}} 
 and it has been adopted as the central hub for knowledge in several domains like 
 life sciences~\cite{BurgstallerMuehlbacher2016}, 
 libraries~\cite{Scott2018} or 
 social science~\cite{Crompton2020}. 
As of August, 2021, it contains information about more than 94 millions of entities~\footnote{\url{https://www.wikidata.org/wiki/Wikidata:Statistics}} 
and since its launch there have been more than 1,400 millions of edits.

\index{Wikibase}
Wikibase~\footnote{\url{https://wikiba.se/}} is a set of 
 open source tools which run Wikidata. 
With Wikibase it is possible to create Knowledge graphs that follow the same data model as Wikidata but that represent information from other domains. 
\index{Wikibase instance}
The projects that are using Wikibase are called Wikibase instances, 
 some examples of wikibase instances are 
 \index{Rhizome}
 Rhizome~\footnote{\url{https://rhizome.org/about/}} or
 \index{Enslaved} Enslaved~\footnote{\url{https://enslaved.org/}}.
 Given that Wikidata was the first and most common Wikibase instance the terms are sometimes used indistinctly. 

\index{Mediawiki}
\index{MariaDB}
Wikibase was initially created from MediaWiki software which ensured adoption by the Wikimedia community. 
Internally, Wikidata content is managed by a relational database (MariaDB) which consists of strings stored and versioned as character blobs~\cite{malyshev2018getting}. 
but was not suitable for advanced data analysis and querying.
With the goal of facilitating those tasks and integrate Wikibase within the semantic web ecosystem, the Wikimedia Foundation adopted 
\index{BlazeGraph}
 BlazeGraph~\footnote{\url{https://blazegraph.com/}} as a complementary triplestore and graph database. 
In this way, there are 2 main data models that coexist in Wikibase: 
 a document-centric model based on MediaWiki and an 
 RDF-based model based on 
 \index{RDF} \index{SPARQL}
 RDF which can be used to do SPARQL queries through the 
 \index{Query service}
 Query Service. 

\begin{figure}
\centering
\includegraphics[width=0.7\textwidth]{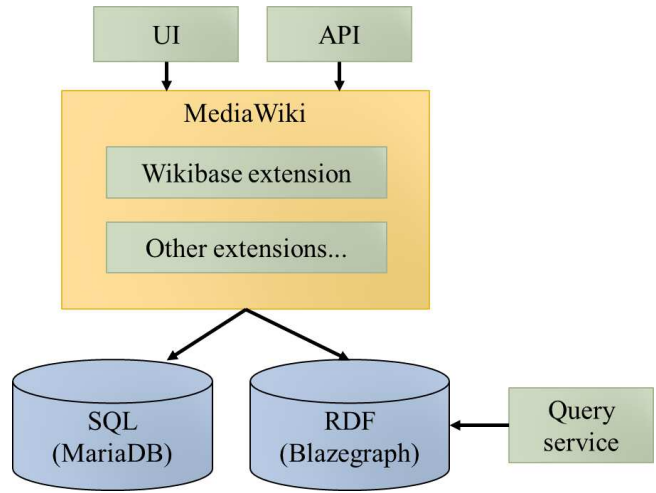}
\caption{Simplified architecture of Wikibase} \label{figure:WikibaseArchitecture}
\end{figure}

A simplified view of Wikibase architecture is depicted in 
figure~\ref{figure:WikibaseArchitecture}~\footnote{A more in-depth view of Wikibase architecture can be found at~\url{https://addshore.com/2018/12/wikidata-architecture-overview-diagrams/}}.

\subsubsection{Wikibase data model: informal introduction}
\index{JSON} \index{RDF}
The Wikibase data model~\footnote{\url{https://www.mediawiki.org/wiki/Wikibase/DataModel}} 
is defined as an abstract data model that can have different serializations like JSON and RDF. 
\index{UML} \index{Wikidata Object Notation}
It is defined using UML data structures and a notation called Wikidata Object Notation. 

\index{Entity} \index{Statement}
Informally, the Wikibase data model is formed from entities and statements about those entities. 
\index{Item}  \index{Property}
 An entity can either be an item or a property. 
 An item is usually represented using a \lstinline|Q| followed by a number and can represent any thing like an abstract of concrete concept. 
  For example, \href{http://www.wikidata.org/entity/Q80}{Q80} 
  represents Tim Berners-Lee in Wikidata.
  
 A property is usually represented by a \lstinline|P| followed by a number and 
  represents a relationship between an item and a value. 
 For example, \href{http://www.wikidata.org/entity/P19}{P19} represents the property \emph{place of birth} in Wikidata.
  
 The values that can be associated to a property are constrained to belong to some specific datatype. 
 There can be compound datatypes like geographical coordinates. 
 
 \index{Datatypes}
 Some of Wikibase datatypes are: quantities, 
  dates and times, 
  geographic locations and shapes, 
  monolingual and 
  multilingual texts, etc. 
 
\index{Statement} 
A statement consists of:

\begin{itemize}
\item A property which is usually denoted using a \lstinline|P| followed by a number.
\item A declaration about the possible value (in wikibase terms, it is called a \emph{snak}) which can be a specific value, \lstinline|no value| declaration or a \lstinline|some value| declaration. 
\item A rank declaration which can be either \lstinline|preferred|, \lstinline|normal| or \lstinline|deprecated|.
\item Zero or more qualifiers which consist of a list of property-value pairs
\item Zero or more references which consist of a list of property-value pairs.
\end{itemize}

\subsubsection{Wikibase data model: formal definition} \label{sec:WikibaseDataModel_FormalDefinition}

We define a formal model for Wikibase which is inspired from 
 Multi-Attributed Relational Structures (MARS)~\cite{MKT2017}. 
For brevity, we model both qualifiers and references as attributes and 
 don't handle the no-value and some-value snaks. 

\begin{definition}[Wikibase graphs] \label{def:wikibaseGraph}
Given a mutually disjoint set of items \ItemSet{}, 
a set of properties \PropSet{} and 
a set of data values \DataValueSet{}, 
a \emph{Wikibase graph} 
  is a tuple $\langle\ItemSet,\PropSet,\DataValueSet,\StmtSet\rangle$ such that
  $\StmtSet\subseteq\EntitySet\times\PropSet\times\ValueSet\times\FinSet{\PropSet\times\ValueSet}$ where 
  $\EntitySet=\ItemSet\cup\PropSet$ is the set of entities 
  which can be subjects of a statement and $\ValueSet=\EntitySet\cup\DataValueSet$ 
  is the set of possible values of a property.
\end{definition}

In practice, Wikibase graphs also add the constraint that 
 every item $q\in\ItemSet$ (or property $p\in\PropSet$) 
 has a unique integer identifier $q^i\in\Nat$ ($p^i\in\Nat$).
 
In the Wikibase data model, statements contain a list of property-values and 
 the values can themselves be nodes from the graph. 
This is different from property graphs, where the set of vertices and the set of values are disjoint. 

\begin{example}[Running example as a Wikibase graph] \label{example:WikibaseGraph}
We continue with the running example about Tim Berners-lee, but extend it with more information about
 awards. 
 More concretely, we add the information that Tim Berners-Lee was awarded with the \emph{Princess of Asturias} ($PA$) award 
 together with Vinton Cerf ($vintCerf$)~\footnote{The award was really obtained by 
  Tim Berners-Lee, Vinton Cerf, Robert Kahn and Lawrence Roberts, 
  we included here only the first two for simplicity}, and that the country of that award is Spain:

\begin{tabular}{ccl}
\ItemSet &= \{ &  \timBl, \vintCerf, \London, \CERN, \UK, \Spain, \PA, \Human \} \\
\PropSet &= \{ &  \birthDate, \birthPlace, \country, \employer, \awarded, \\
 & & \start, \pEnd, \pointTime, \togetherWith, \instanceOf \} \\
\DataValueSet &= \{ & 1984,1994,1980,1955\} \\
$\StmtSet$ &= \{ & (\timBl, \instanceOf, \Human, \{\}), \\ 
 & & (\timBl, \birthDate, 1955, \{\}), \\ 
 & & (\timBl, \birthPlace, \London, \{\}), \\
 & & (\timBl, \employer, \CERN, \{ \start:1980, \pEnd:1980 \}), \\
 & & (\timBl, \employer, \CERN, \{ \start:1984, \pEnd:1994 \}), \\
 & & (\timBl, \awarded, \PA, \{\pointTime: 2002, \togetherWith:\vintCerf\}), \\
 & & (\London, \country, \UK, \{\}), \\
 & & (\vintCerf, \instanceOf, \Human, \{\}) \\
 & & (\vintCerf, \birthPlace, \newHaven, \{\}) \\
 & & (\CERN, \awarded, \PA, \{ \pointTime: 2013 \}) \\
 & & (\PA, \country, \Spain, \{ \}) \} \\
\end{tabular}

Figure~\ref{figure:WikibaseGraphExample} presents a possible visualization of a wikibase graph.

\begin{figure}
\centering
\includegraphics[width=\textwidth]{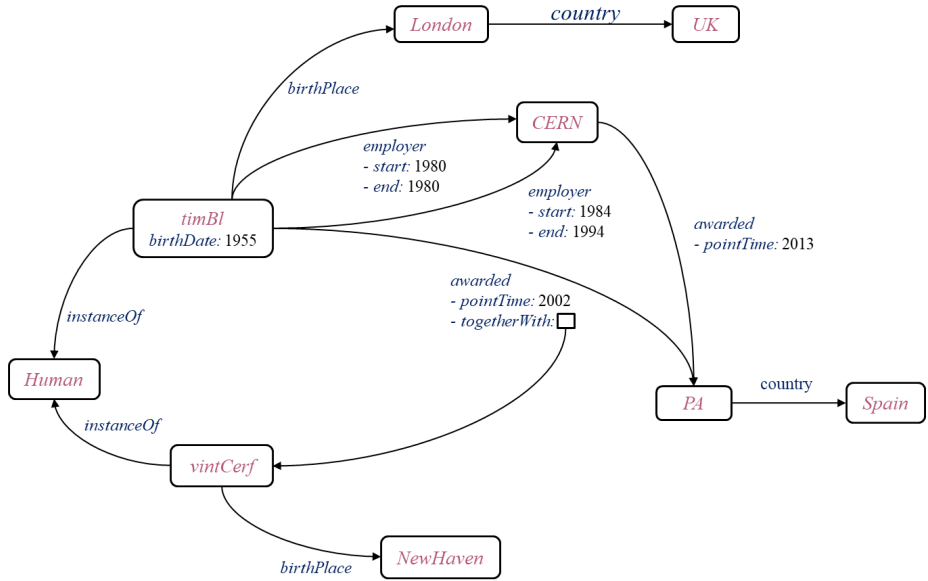}
\caption{Visualization of example wikibase graph} \label{figure:WikibaseGraphExample}
\end{figure}

\end{example}

The Wikibase data model supports 2 main export formats: JSON and RDF.
 The \indexn{JSON} one directly follows the structure of the Wikibase data model and 
  is employed by the JSON Dumps 
  while the \indexn{RDF serialization} follows semantic web and linked data principles. 

\subsubsection{Wikibase JSON serialization} \label{WikibaseJSONSerialization}

The JSON serialization follows the Wikibase data model. It basically consists of an array of entities 
 where each entity is a JSON object that captures all the local information about the entity: 
  the labels, descriptions, aliases, sitelinks and statements that have the entity as subject.
  Each JSON object is represented in a single line.
A remarkable feature of this encoding is that it captures the output neighborhood of 
 every entity in a single line making it amenable to processing models that focus on local neighborhoods
 because the whole graph can be processed in a single pass.

\begin{lstlisting}[basicstyle=tiny, style=json]
[
 { "type": "item", "id": "Q42", "claims": { "P31": [...
 { "type": "item", "id": "Q80", "claims": { "P108": [...
 { "type": "property", "id": "P108", "claims": { ... 
 ... 
]
\end{lstlisting}

%

\subsubsection{Wikibase RDF serialization}

The RDF serialization\footnote{\url{https://www.mediawiki.org/wiki/Wikibase/Indexing/RDF_Dump_Format}} of 
 Wikidata was designed with the goal of being able to represent all the structures of the Wikibase data model in RDF, maintaining compatibility with semantic web vocabularies like RDFS and OWL and avoiding the use of blank nodes~\cite{EGKMV2014}. 

\begin{example}[RDF serialization of a node]
As an example, the information about Tim Berners-Lee (Q80) declaring that he was as employer (P108) of 
 CERN (Q42944) between 1984 and 1994 is represented as~\footnote{The full Turtle serialization can be obtained at:~\url{https://www.wikidata.org/wiki/Special:EntityData/Q80.ttl}}:

\begin{lstlisting}[style=Turtle]
wd:Q80 rdf:type wikibase:Item ;
 wdt:P108 wd:Q42944     ;
 p:P108 :Q80-4fe7940f   .
       
:Q80-4fe7940f rdf:type wikibase:Statement ;
 wikibase:rank wikibase:NormalRank ;
 ps:P108       wd:Q42944 ;
 pq:P580       "1984-01-01T00:00:00Z"^^xsd:dateTime ;
 pq:P582       "1994-01-01T00:00:00Z"^^xsd:dateTime .
\end{lstlisting}

The RDF serialization uses a direct arc to represent the preferred statement 
 represented by prefix alias \lstinline|wdt:| 
 leaving the rest of the values of a property accessible through the namespaces 
  \lstinline|p:|, \lstinline|ps:| and \lstinline|pq:|.

The reification model employed by Wikidata creates auxiliary nodes that represent each statement. In the previous example, the node \lstinline|:Q80-4fe7940f| represents the statement which can be qualified with the start and end time. 

\end{example}

Apart of the the dumps, RDF serialization is also employed by the 
 Wikidata Query Service~\cite{BGK2018,MKGGB2018} and users of 
 Wikidata are required to use and understand the singleton statement approach and namespace conventions 
 employed. 


%
%

%% file: 30_DescribingKGs.tex
\section{Describing Knowledge Graphs} \label{sec:DescribingKGs}

\subsection{Describing and validating RDF} \label{sec:ShEx}

At the end of 2013, an \emph{RDF Validation Workshop}~\footnote{\url{https://www.w3.org/2012/12/rdf-val/}} was organized by W3C/MIT to discuss use cases and requirements related with the quality of RDF data. 
One of the conclusions of the workshop was that there was a need for a high-level language that could describe and validate RDF data.

Shape Expressions (ShEx) were proposed as such a language in 2014~\cite{Prudhommeaux2014}. 
It was designed as a high-level and concise domain-specific language to describe RDF. The syntax of ShEx is inspired by Turtle and SPARQL, while the semantics is inspired by RelaxNG and XML Schema.

In this section we describe a simplified abstract syntax of ShEx following~\cite{Boneva2017}\footnote{The full specification of ShEx is available at~\url{https://shex.io/shex-semantics/}}.

\begin{definition}[ShEx schema] \label{def:ShExSchema}
A \indexe{ShEx Schema} is defined as a tuple $\langle\LabelSet,\schemaDef\rangle$ 
where 
$\LabelSet$ set of shape labels, 
and $\schemaDef : \LabelSet\rightarrow\ShapeSet$ is a total function from labels to shape expressions.

The set of shape expressions $se\in\ShapeSet$ is defined using the following abstract syntax:

\begin{tabular}{ccll}
  $se$   & ::= &  cond  & Basic boolean condition on nodes (node constraint)\\
         & $|$ & $s$ & Shape \\
         & $|$ & $se_1$ \c|AND| $se_2$  & Conjunction \\
         & $|$ & @\lbl & Shape label reference for $\lbl\in\LabelSet$ \\
  $s$    & ::= & \c|CLOSED| $\{ te \}$ & Closed shape \\
         & $|$ & $\{ te \}$ & Open shape \\
  $te$   & ::= & \eachOf{te_1}{te_2} & Each of $te_1$ and $te_2$ \\
         & $|$ & \oneOf{te_1}{te_2} & Some of $te_1$ or $te_2$ \\
         & $|$ & $te*$ & Zero or more ${te}$  \\
         & $|$ & $\epsilon$  & Empty triple expression \\
         & $|$ & \arc{p}{@\lbl} & Triple constraint with predicate $p$ \\
\end{tabular}

\end{definition}

Intuitively, shape expressions define conditions about nodes
 while triple expressions define conditions about the neighborhood of nodes, 
 and shapes qualify those neighborhoods by disallowing triples with other predicates 
 in the case of closed shapes or allowing them in the case of open shapes.

In this paper we omit the negation and disjunction operator to facilitate the presentation of the subsetting semantics. 
 Adding those operators increases the expressiveness of ShEx to validate RDF graphs but 
  we consider that their use to create subsets is not yet clear so we decided to leave them for further research.
 
The restrictions imposed on shape expressions schemas in~\cite{ShExSpec} also apply here. 
Namely, in a schema $(\LabelSet,\schemaDef,\ShapeSet)$

\begin{itemize}
\item The shape label references used by the definition function $\schemaDef$ are themselves defined, i.e. 
if $@\lbl$ appears in some shape definition, then $\lbl$ belongs to $\LabelSet$;
\item No definition $\schemaDef(\lbl)$ uses a reference $@\lbl$ to itself,
 neither directly nor transitively, except while traversing a shape. 
 For instance, $\schemaDef(\lbl) = @\lbl\,\c|AND|\; se$ is forbidden, 
 but $\schemaDef(\lbl) = \{ \arc{p}{@\lbl} \}$ is allowed.
\end{itemize}

\begin{example}[Example of ShEx schema]

A ShEx schema that describes the RDF graph presented in example~\ref{example:RDFData} can be defined as:

\begin{tabular}{ccll}
\LabelSet & = & \{ & $Person$,$Place$,$Country$,$Organization$,$Date$\} \\
\schemaDef($Person$) & = & \{ & \arc{birthDate}{@Date}; \arc{birthPlace}{@Place}; \\
                     &   &    & \arc{employer}{@Organization\,*} \} \\
\schemaDef($Place$)  & = & \{ & \arc{country}{@Country}\} \\
\schemaDef($Country$)  & = & \{ & \} \\
\schemaDef($Organization$)  & = & \{ & \} \\ 
\schemaDef($Date$)  & = & & xsd:Date \\ 
\end{tabular}
\end{example}

ShEx has several concrete syntaxes like a compact syntax (ShExC) and an RDF syntax defined based on JSON-LD (ShExJ)~\footnote{See~\cite{ShExSpec} for details.}. 

\begin{example}[Example of ShEx in ShExC compact syntax]

The previous ShEx schema can be defined using the compact syntax as:

\begin{lstlisting}[style=ShExC]
:Person {
  :birthPlace @:Place ;
  :birthDate  @:Date ;
  :employer   @:Organization ;
}
:Place {
  :country    @:Country
}
:Country      {}
:Organization {}
:Date         {}
\end{lstlisting}

In general, it is possible to visualize ShEx schemas using UML-like class diagrams. 
Figure~\ref{fig:ShExVisualization} presents a visualization of the previous schema using RDFShape~\footnote{This visualization can be interactively generated following:~\url{https://rdfshape.weso.es/link/16344153229}}

\begin{figure}
\centering
\includegraphics[width=0.7\textwidth]{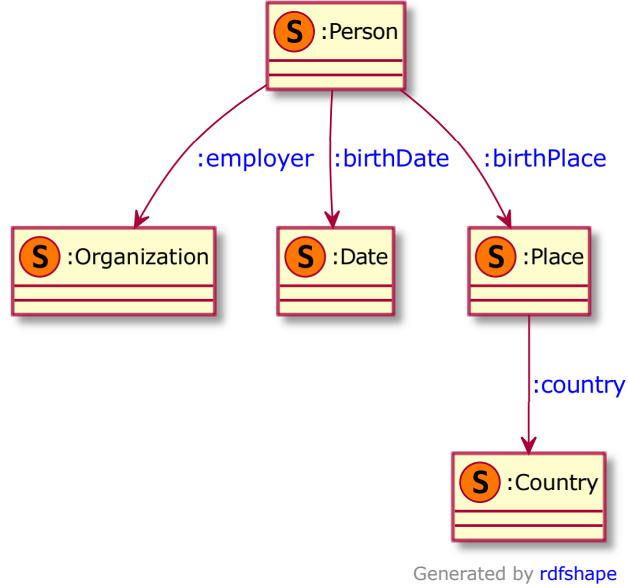}
\caption{ShEx schema visualization as UML-like diagrams} \label{fig:ShExVisualization}
\end{figure} 

\end{example}

Apart of describing RDF data, Shape Expressions have been designed to enable validation and 
 checking if an RDF node conforms to some shape. 

The semantics of Shape Expression validation can be defined with a relation 
 between an RDF node, an RDF graph, a ShEx schema and a shape assignment. 

As an example of validation, we have implemented the ShEx-s library which is used by 
 RDFShape~\footnote{It is possible to see the results of validating the previous example 
  in RDFShape following this link: https://rdfshape.weso.es/link/16275436158}.

\label{ShExSemantics}
The semantics of ShEx schemas is based on a conformance relation parameterized by a shape assignment: 
 we say that node $n$ in graph $\Graph$ conforms to shape expression $se$ with shape assignment $\typing$, 
 and we write $\conforms{\Graph}{n}{\typing}{se}$.

The following rules are defined similar to~\cite{Boneva17}, where it is shown that there exists a unique maximal shape assignment $\typing_{\text{max}}$ that allows to define conformance independently on the shape assignment.

The conformance relation is defined recursively on the structure of $se$ by the set of inference rules presented in table~\ref{sem:ShEx_SE} where $\props{te}$ is the set of predicates that appear in a triple expression $te$ and can be defined as:

\begin{table}[h!]
\begin{tabular}{c}
\inference[$Cond$]
  {cond(n)=true} 
  {\conforms{\Graph}{n}{\typing}{cond}}

  \hspace{1cm}
\inference[$AND$]
  {\conforms{\Graph}{n}{\typing}{se_1} & 
   \conforms{\Graph}{n}{\typing}{se_2}
  } 
  {\conforms{\Graph}{n}{\typing}{se_1 \text{ \texttt{AND} } se_2}}
\\ \\
  
%
%
%

  \inference[$ClosedShape$]
  { neighs(n,\Graph) = ts & 
   \conformsTE{\Graph}{ts}{\typing}{te}
  } 
  {\conforms{\Graph}{n}{\typing}{\text{\texttt{CLOSED} }\{te\}}}  
  \\ \\

  \inference[$OpenShape$]
  { ts = \{\triple{x}{p}{y} \in neighs(n,\Graph) \mid p \in \props{te}\} &
    \conformsTE{\Graph}{ts}{\typing}{te}
  } 
  {\conforms{\Graph}{n}{\typing}{\{te\}}}
\end{tabular}
\caption{Inference rules for ShEx shape expressions} \label{sem:ShEx_SE}
\end{table}
 
\begin{tabular}{lll}
$\props{\eachOf{te_1}{te_2}}$ & = & $\props{te_1}\cup\props{te_2}$ \\
$\props{\oneOf{te_1}{te_2}}$ & = & $\props{te_1}\cup\props{te_2}$ \\
$\props{\arc{p}{te}}$ & = & $\{p\}$ \\
$\props{te*}$ & = & $\props{te}$ \\
$\props{\epsilon}$ & = & $\emptyset$ \\
\end{tabular}

The rules for node constraint ($Cond$) and conjunction are as expected.
A node $n$ conforms to an open shape with triple expression $te$ 
 if its neighborhood restricted to the triples with predicates from $te$ conform, 
 meaning that triples whose predicates are not mentioned in $te$ are not constrained by the shape (rule $OpenShape$).
Conformance to a closed shape requires to consider the whole neighborhood 
 of the node (rule $ClosedShape$).

Conformance to a triple expression uses a second conformance relation defined on sets on neighborhood triples $ts$ instead of nodes $n$.
 The set of neighborhood nodes $ts$ of a graph $\Graph$ conforms to a triple expression $te$ with shape assignment $\typing$, 
  written as $\conformsTE{\Graph}{ts}{\typing}{te}$, 
  as defined by the inference rules in table~\ref{sem:ShEx_TE}.

\begin{table}[h!]
\begin{tabular}{c}
\\ \\
\inference[$EachOf$]
  { (ts_1,ts_2)\in\partition{ts} & 
    \conformsTE{\Graph}{ts_1}{\typing}{te_1} &
    \conformsTE{\Graph}{ts_2}{\typing}{te_2} 
  }
  {\conformsTE{\Graph}{ts}{\typing}{\eachOf{te_1}{te_2}}}

\\ \\

\inference[$OneOf_1$]
  { \conformsTE{\Graph}{ts}{\typing}{te_1} }
  {\conformsTE{\Graph}{ts}{\typing}{\oneOf{te_1}{te_2}} }

\hspace{1cm} 
\inference[$OneOf_2$]
  { \conformsTE{\Graph}{ts}{\typing}{te_2} }
  {\conformsTE{\Graph}{ts}{\typing}{\oneOf{te_1}{te_2}}}

\\ \\
\inference[$TripleConstraint$]
  {ts = \{\triple{x}{p}{y}\} & \conforms{\Graph}{y}{\typing}{@\lbl} }
  {\conformsTE{\Graph}{ts}{\typing}{\arc{p}{@\lbl}}}

\hspace{0.5cm}
\inference[$Star_1$]
  {}
  {\conformsTE{\Graph}{\emptyGraph}{\typing}{te*}}
\\ \\
\inference[$Star_2$]
  {(ts_1,ts_2)\in\partition{ts} & 
    \conformsTE{\Graph}{ts_1}{\typing}{te} &
    \conformsTE{\Graph}{ts_2}{\typing}{te*}}
  {\conformsTE{\Graph}{ts}{\typing}{te*}}
\end{tabular}
\caption{Inference rules for ShEx triple expressions} \label{sem:ShEx_TE}
\end{table} 

The semantics of ShEx schema can be defined independently of shape assignments.
A shape assignment $\typing$ for graph $\Graph$ and $\Schema$ is called \emph{valid} 
if for every node $n$ in $\Graph$ and every shape expression label $\lbl$ defined in $\Schema$, 
if $\hasType{n}{\lbl} \in \typing$, then $\conforms{\Graph}{n}{\typing}{@\lbl}$.
  
\begin{lemma}[Boneva et al~\cite{Boneva2017}]
  For every graph $\Graph$, there exists a unique maximal valid shape shape assignment $\typing_{\text{max}}$ such that if $\typing$ is a valid shape assignment for $\Graph$ and $\Schema$, 
  then $\typing \subseteq \typing_{\text{max}}$.
\end{lemma}


\subsection{Describing and validating Property graphs}

In this section we define a ShEx extension called PShEx that can be used to describe and validate Property graphs.
 According to the definition of property graphs given in section~\ref{sec:PropertyGraphs}, 
 nodes and edges can have associated labels as well as a set of property/values. 
 In this way, it is necessary to adapt the definition of ShEx to describe pairs or property/values that we will call qualifiers.
 
The language PShEx is composed of three main categories: 
 shape expressions ($se$) that describe the shape of nodes, 
 triple expressions ($te$) that describe the shape of edge relationships and 
 qualifier expressions ($qs$) that describe qualifiers sets of property/values associated with node/edge identifiers.

\begin{definition}[PShEx schema]
A \indexe{PShEx Schema} is a tuple $\langle\LabelSet,\schemaDef\rangle$ 
where 
$\LabelSet$ set of shape labels, 
and $\schemaDef : \LabelSet\rightarrow\ShapeSet$ is a total function from labels to shape expressions $se\in\ShapeSet$ defined using the abstract syntax:

\begin{tabular}{ccll}
$se$   & ::= &  $cond_{t_s}$  & Basic boolean condition on set of types $t_s\subseteq\PLabelSet$\\
 & $|$ & $s$ & Shape \\
 & $|$ & $se_1$ \c|AND| $se_2$  & Conjunction \\
 & $|$ & @\lbl & Shape label reference for $\lbl\in\LabelSet$ \\
 & $|$ & qs & Qualifiers of that node \\
$s$    & ::= & \c|CLOSED| $\{ te \}$ & Closed shape \\
 & $|$ & $\{ te \}$ & Open shape \\
$te$   & ::= & \eachOf{te_1}{te_2} & Each of $te_1$ and $te_2$ \\
 & $|$ & \oneOf{te_1}{te_2} & Some of $te_1$ or $te_2$ \\
 & $|$ & $te*$ & Zero or more ${te}$  \\
 & $|$ & \arc{p}{@\lbl\,\,qs} & Triple constraint with property type $p$ \\
 &     &  & whose nodes satisfy the shape $\lbl$ and qualifiers $qs$ \\
$qs$ & ::= & \openQs{ps} & Open qualifier specifiers $ps$ \\
  & $|$ & \closeQs{ps}   & Closed qualifier specifiers $ps$ \\
$ps$ & ::= & $\eachOfQs{ps_1}{ps_2}$ & Each of $ps_1$ and $ps_2$ \\
 & $|$ & $\oneOfQs{ps_1}{ps_2}$  & OneOf of $ps_1$ or $ps_2$ \\
 & $|$ & $ps*$ & zero of more $ps$ \\
 & $|$ & $p:cond_{v}$ & Property $p$ with value conforming to $cond_{v}$ \\
 & & & $cond_{v_s}$ is a boolean condition on sets of values $v_s\subseteq\ValueSet$ \\
\end{tabular}

\end{definition}

We will omit the list of qualifiers when it is empty.

\begin{example}\label{example:PropertyGraphsSchema}

As an example, we can define a PShEx schema that describes the property graph from example~\ref{example:PropertyGraphs} 
where $\condType{t}$ is a condition that is satisfied when the 
 set of types of a node contains the type $t$, i.e. 
$\condType{t}(vs)=\text{\c|true| if }t\in{}vs$ and 
$String, Date$ are conditions on the values that are satisfied when
 the values have the corresponding type.
	
\begin{tabular}{ccll}
	\LabelSet & = & \{ & $Person, Place, Country, Org $\} \\
\schemaDef(Person) & = & & \condType{Human}\,\c|AND|\,\openQs{label:String, birthDate:Date}\,\c|AND|\,\{ \\ 
      & & & \arc{birthPlace}{@Place}; \\
      & & & \arc{employer}{@Org\,\,\openQs{start:Date, end:Date}}* \\
      & & \} & \\
\schemaDef(Place) &= & & \openQs{label:String}\,\c| AND |\,\{ \\ 
  & & & \arc{country}{@Country} \\
  & & \} & \\
\schemaDef(Country) & = & & \condType{Country}\,\c|AND|\,\openQs{label:String} \{\} \\
\schemaDef(Org) & = & & \condType{Organization}\,\c|AND|\,\openQs{label:String} \{\} \\
\end{tabular}

\end{example}

In order to define the semantic specification of PShEx we will need to define the neighborhood of a node in a property graph.

\index{Neighbourhood of node in property graph}
\begin{definition}[Neighborhood of node in property graph]
The neighbors of a node $n\in\NodeSet$ in a property graph $\Graph=\langle\NodeSet,\EdgeSet,\rho,\lambda_n,\lambda_e,\sigma\rangle$ are defined as 
$neighs(n)=\{(n,p,y,vs) \mid \exists{}v\in\EdgeSet\text{ such that }\rho(v)=(n,y)\wedge\lambda_e(v)=p\wedge{}vs=\{(k,v)\mid{}\sigma(k,v)=ws\wedge{}v\in{}ws\}\}$
\end{definition}

\begin{example}
The neighbors of node $n_1$ in property graph~\ref{example:PropertyGraphs} are:

\begin{tabular}{ccll}
$neighs(n_1)$ & = $\{$ & $(n_1,birthPlace,n_2,\{\}),$ \\
 & & $(n_1,employer,n_4,\{(start,1980),(end,1980)\}),$ \\
 & & $(n_1,employer,n_4,\{(start,1984),(end,1994)\}) \} $
\end{tabular}
\end{example}	

The semantic specification of PShEx can defined in a similar way to the ShEx one.
 Given a property graph $\Graph$, 
  and a shape assignment $\typing$, a node identifier $n\in\NodeSet$ conforms with a 
  shape expression $se$, which is represented as $\conforms{\Graph}{n}{\typing}{se}$ 
  and follows the rules presented in~\ref{def:SE_PShEx} where $\props{te}$ is the set of edge labels (or predicates) 
  that appear in a triple expression $te$ and can be defined as:

\begin{table}[h!]
\begin{tabular}{c}
	\inference[$Cond_{ts}$]
	{\lambda_n(n)=vs & cond_{ts}(vs)=true} 
	{\conforms{\Graph}{n}{\typing}{cond_{ts}}}
	
	\hspace{1cm}
	\inference[$AND$]
	{\conforms{\Graph}{n}{\typing}{se_1} & 
		\conforms{\Graph}{n}{\typing}{se_2}
	} 
	{\conforms{\Graph}{n}{\typing}{se_1 \text{ \texttt{AND} } se_2}}
	\\ \\
	
%
%
%
	
	\inference[$ClosedShape$]
	{ neighs(n,\Graph) = ts & 
		\conformsTE{\Graph}{ts}{\typing}{s'}
	} 
	{\conforms{\Graph}{n}{\typing}{\text{\texttt{CLOSED} }\{te\}}}  
	\\ \\
	
	\inference[$OpenShape$]
	{ ts = \{\triple{x}{p}{y} \in neighs(n,\Graph) \mid p \in \props{te}\} &
		\conformsTE{\Graph}{ts}{\typing}{te}
	} 
	{\conforms{\Graph}{n}{\typing}{\{te\}}}
\end{tabular}
\caption{Rules for PShEx shape expressions} \label{def:SE_PShEx}
\end{table}

\begin{tabular}{lll}
	$\props{\eachOf{te_1}{te_2}}$ & = & $\props{te_1}\cup\props{te_2}$ \\
	$\props{\oneOf{te_1}{te_2}}$ & = & $\props{te_1}\cup\props{te_2}$ \\
	$\props{\arc{p}{te}}$ & = & $\{p\}$ \\
	$\props{te*}$ & = & $\props{te}$ \\
	$\props{\epsilon}$ & = & $\emptyset$ \\
\end{tabular}

As in the case of ShEx, the previous definition uses a second conformance 
 relation defined on sets of triples $ts$ instead of nodes $n$.
 The set of neighborhood nodes $ts$ from a property graph $\Graph$ conforms to a triple expression $te$ 
 with shape assignment $\typing$, written $\conformsTE{\Graph}{ts}{\typing}{s}$, 
 as defined by the inference rules represented in table~\ref{sem:PShEx_TE}.

\begin{table}[h!]
\begin{tabular}{c}
	\\
	\inference[$EachOf$]
	{ (ts_1,ts_2)\in\partition{ts} & 
		\conformsTE{\Graph}{ts_1}{\typing}{te_1} &
		\conformsTE{\Graph}{ts_2}{\typing}{te_2} 
	}
	{\conformsTE{\Graph}{ts}{\typing}{\eachOf{te_1}{te_2}}}
	
	\\ \\
	
	\inference[$OneOf_1$]
	{ \conformsTE{\Graph}{ts}{\typing}{te_1} }
	{\conformsTE{\Graph}{ts}{\typing}{\oneOf{te_1}{te_2}} }
	
	\hspace{1cm} 
	\inference[$OneOf_2$]
	{ \conformsTE{\Graph}{ts}{\typing}{te_2} }
	{\conformsTE{\Graph}{ts}{\typing}{\oneOf{te_1}{te_2}}}
	
	\\ \\
	
	\inference[$TripleConstraint$]
	{ts = \{\quadruple{x}{p}{y}{s}\} & 
		\conforms{\Graph}{y}{\typing}{@\lbl} &
		\conformsQs{\Graph}{s}{\typing}{qs}
	}
	{\conformsTE{\Graph}{ts}{\typing}{\arc{p}{@\lbl}\,\,qs}}
	
	\\ \\

	\inference[$Star_1$]
	{}
	{\conformsTE{\Graph}{\emptyGraph}{\typing}{te*}}

	\\ \\
	\inference[$Star_2$]
	{(ts_1,ts_2)\in\partition{ts} & 
		\conformsTE{\Graph}{ts_1}{\typing}{te} &
		\conformsTE{\Graph}{ts_2}{\typing}{te*}}
	{\conformsTE{\Graph}{ts}{\typing}{te*}}
	\\ \\
\end{tabular}
\caption{Rules for PShEx triple expressions} \label{sem:PShEx_TE}
\end{table}

In the case of PShEx we declare a new conformance relationship $\conformsQs{\Graph}{s}{\typing}{qs}$ 
between a graph $\Graph$ 
a set $s\in{}P\times{}V$ of property-value elements, 
a shape assignment $\typing$ and 
a qualifier specifier $qs$ whose rules are defined in table~\ref{sem:PShEx_Qs}
where $props(ps)$ is the set of properties that appear in a property specifier $ps$ and can be defined as:

\begin{table}[h!]
\begin{tabular}{c}
	\hspace{-0.8cm}\inference[$OpenQs$]
	{s'=\{(p,v)\in{}s|p\in{}props(ps)\} &
		\conformsQs{\Graph}{s'}{\typing}{ps}
	}
	{\conformsQs{\Graph}{s}{\typing}{\openQs{ps}}} 
	
	\hspace{0.4cm}
	
	\inference[$CloseQs$]
	{\conformsQs{\Graph}{s}{\typing}{ps}}
	{\conformsQs{\Graph}{s}{\typing}{\closeQs{ps}}} 
	
	\\ \\
	
	\inference[$EachOfQs$]
	{\conformsQs{\Graph}{s}{\typing}{ps_1} &
		\conformsQs{\Graph}{s}{\typing}{ps_2}
	}
	{\conformsQs{\Graph}{s}{\typing}{\eachOfQs{ps_1}{ps_2}}} 
	
	\\ \\
	
	\inference[$OneOfQs_1$]
	{\conformsQs{\Graph}{s}{\typing}{ps_1}}
	{\conformsQs{\Graph}{s}{\typing}{\oneOfQs{ps_1}{ps_2}}} 
	
	\hspace{0.5cm}
	
	\inference[$OneOfQs_2$]
	{\conformsQs{\Graph}{s}{\typing}{ps_2} }
	{\conformsQs{\Graph}{s}{\typing}{\oneOfQs{ps_1}{ps_2}}} 
	
	\\ \\
	\hspace{-0.8cm}
	\inference[$StarQs_1$]
	{}
	{\conformsQs{\Graph}{\emptyset}{\typing}{ps*}} 
	
	\hspace{0.4cm}
	
	\inference[$StarQs_2$]
	{(s_1,s_2)\in{}\partition{s} & 
		\conformsQs{\Graph}{s_1}{\typing}{ps} &
		\conformsQs{\Graph}{s_2}{\typing}{ps*}
	}
	{\conformsQs{\Graph}{s}{\typing}{ps*}} 
	
	\\ \\
	
	\inference[$PropertyQs$]
	{ s = \{(p,w)\} &
		conv_v(w)=\c|true|
	}
	{\conformsQs{\Graph}{s}{\typing}{p:cond_v}} 
	
\end{tabular}
\caption{Rules for PShEx qualifiers} \label{sem:PShEx_Qs}
\end{table}

\begin{tabular}{lll}
	$props(\eachOfQs{ps_1}{ps_2})$ & = & $props(ps_1)\cup{}props(ps_2)$ \\
	$props(\oneOfQs{ps_1}{ps_2})$ & = & $props(ps_1)\cup{}props(ps_2)$ \\
	$props(ps*)$ & = & $\props{ps}$ \\
	$props(p:cond_v)$ & = & $\{p\}$ \\
\end{tabular}

As in the case of ShEx, the semantics of ShEx schemas can be defined independently on shape assignments.
A shape assignment $\typing$ for graph $\Graph$ and $\Schema$ is called \emph{valid} 
if for every node $n$ in $\Graph$ and every shape expression label $\lbl$ defined in $\Schema$, 
if $\hasType{n}{\lbl} \in \typing$, then $\conforms{\Graph}{n}{\typing}{@\lbl}$.

\subsection{Describing and validating Wikibase graphs} \label{sec:WShEx}

Wikidata adopted ShEx in 2019 as the language to define entity schemas which can be used to validate entities. Nevertheless, they describe the RDF serialization of Wikibase entities instead of the Wikibase datamodel. 
This requires users to be aware of how qualifiers and references are serialized in Wikibase which can lead to duplicated properties. 
Another problem of ShEx schemas is that they cannot be used to directly describe the contents of Wikidata dumps in JSON which are closer to the Wikibase data model. 

\begin{figure}
\centering
\includegraphics[width=0.7\textwidth]{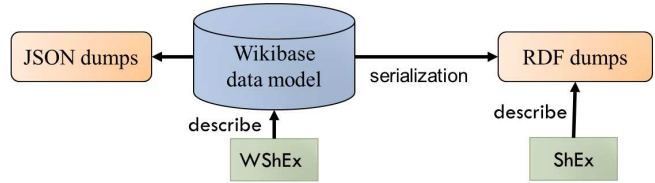}
\caption{Relationship between ShEx, WShEx and Wikibase data model} \label{fig:ShEx_vs_WShEx}
\end{figure}

To that end, we designed an extension of ShEx called WShEx that can describe the Wikibase data model and so, be used to validate Wikibase dumps in JSON without requiring them to be serialized in RDF.
Figure~\ref{fig:ShEx_vs_WShEx} represents the relationship between ShEx and WShEx.

WShEx is presented as an extension of the ShEx language defined in section~\ref{sec:ShEx} 
adapted to the wikibase graphs definitions~\ref{def:wikibaseGraph}.

\begin{definition}[WShEx schema]
A \emph{WShEx Schema} is defined as a tuple $\langle\LabelSet,\schemaDef\rangle$ 
where 
$\LabelSet$ set of shape labels, 
and $\schemaDef : \LabelSet\rightarrow\ShapeSet$ is a total function from labels to w-shape expressions.

\end{definition}

The set of shape expressions $se\in\ShapeSet$ is defined using the abstract syntax presented in table~\ref{table:abstractSyntaxShEx}. 
Notice that it is an extension of the abstract syntax for ShEx modifying the rule for triple constraint 
adding a new element for qualifier specifiers and 
adding the corresponding rules for qualifier specifiers.

\begin{table}
\begin{tabular}{ccll}
  $se$   & ::= &  cond  & Basic boolean condition on nodes (node constraint)\\
         & $|$ & $s$ & Shape \\
         & $|$ & $se_1$ \c|AND| $se_2$  & Conjunction \\
         & $|$ & @\lbl & Shape label reference for $\lbl\in\LabelSet$ \\
  $s$    & ::= & \c|CLOSED| $s'$ & Closed shape \\
         & $|$ & $s'$ & Open shape \\
  $s'$   & ::= & \{ te \} & Shape definition \\
  $te$ & ::= & \eachOf{te_1}{te_2} & Each of $te_1$ and $te_2$ \\
         & $|$ & \oneOf{te_1}{te_2} & Some of $te_1$ or $te_2$ \\
         & $|$ & $te*$ & Zero or more ${te}$  \\
         & $|$ & \arc{p}{@\lbl\,\,qs} & Triple constraint with predicate $p$ \\
         & & & value conforming to $\lbl$ and qualifier specifier $qs$ \\
         & $|$ & $\epsilon$  & Empty triple expression\\
  $qs$ & ::= & \openQs{ps}  & Open property specifier \\
       & $|$ & \closeQs{ps}      & Closed property specifier \\
  $ps$ & ::= & \eachOfQs{ps}{ps} & \emph{EachOf} property specifiers \\
       & $|$ & \oneOfQs{ps}{ps}  & \emph{OneOf} property specifiers \\
       & $|$ & ps* & zero of more property specifiers \\
       & $|$ & $\epsilon$ & Empty property specifier \\
       & $|$ & p:@\lbl & Property $p$ with value conforming to shape $\lbl$ \\
\end{tabular}
\caption{Abstract syntax of WShEx} \label{table:abstractSyntaxShEx}
\end{table}
\begin{example}[Example of WShEx schema]

A ShEx schema that describes the Wikibase graph presented in example~\ref{example:WikibaseGraph} can be defined as:

\begin{tabular}{ccll}
\LabelSet & = & \{ & $Person$,$Place$,$Country$,$Organization$,$Date$, $Award$\} \\
\schemaDef($Person$) & = & \{ & \arc{birthDate}{@Date}; \arc{birthPlace}{@Place}; \\
                     &   &  & \arc{employer}{@Organization} \openQs{\eachOfQs{start:@Date}{end:@Date}}* \\
                     &   &  & \arc{awarded}{@Award} \openQs{\eachOfQs{pointTime:@Date}{togetherWith:@Person}}* \\
                     & & \} & \\
\schemaDef($Place$)  & = & \{ & \arc{country}{@Country}\} \\
\schemaDef($Country$)  & = & \{ & \} \\
\schemaDef($Award$)  & = & \{ & \arc{country}{@Country}\} \\
\schemaDef($Organization$)  & = & \{ & \} \\
\schemaDef(Date)  & = &  & $\in{}xsd:date$\\
\end{tabular}
\end{example}

It is possible to define a compact syntax for WShEx in a similar way to ShExC adding the symbols \lstinline|{{...}}| 
 to declare open qualifier specifiers and 
 \lstinline|[[...]]| for closed ones.

\begin{example}[Example of WShEx schema using the compact Syntax]
\begin{lstlisting}[style=ShExC]
:Researcher {
 birthPlace     @<Place>          ;     
 birthDate      @<Time>           ; 
 employer       @<Organization> * 
    {{ :start   @:Date, 
       :end     @:Date
    }} ;
 awarded       @<Award> * 
    {{ :pointTime    @:Date, 
       :togetherWith @:Person
    }} 
}
:Place          { country @<Country> }
:Organization   {}
:Award          { country @<Country> }
:Country        {}
:Date           xsd:date
\end{lstlisting}
\end{example}

The semantics of WShEx is similar to the semantics defined for  ShEx and PShEx. 
We define a conformance relation parameterized by a shape assignment $\conforms{\Graph}{n}{\typing}{se}$ with the meaning that node $n$ in graph $\Graph$ conforms to shape expression $se$ with shape assignment $\typing$ according to the rules~\ref{sem:WShEx_SE}.

\begin{table}[h!]
\begin{tabular}{c}
\inference[$Cond$]
  {cond(n)=true} 
  {\conforms{\Graph}{n}{\typing}{cond}}

  \hspace{1cm}
\inference[$AND$]
  {\conforms{\Graph}{n}{\typing}{se_1} & 
   \conforms{\Graph}{n}{\typing}{se_2}
  } 
  {\conforms{\Graph}{n}{\typing}{se_1 \text{ \texttt{AND} } se_2}}
\\ \\
  
%
%

  \inference[$ClosedShape$]
  { neighs(n,\Graph) = ts & 
   \conformsTE{\Graph}{ts}{\typing}{s'}
  } 
  {\conforms{\Graph}{n}{\typing}{$\c|CLOSED|\,$s'}}  
  \\ \\

  \inference[$OpenShape$]
  { ts = \{\triple{x}{p}{y} \in neighs(n,\Graph) \mid p \in \props{te}\} &
    \conformsTE{\Graph}{ts}{\typing}{s'}
  } 
  {\conforms{\Graph}{n}{\typing}{s'}}
  \\ \\
\end{tabular}
\caption{Inference rules for WShEx shape expressions} \label{sem:WShEx_SE}
\end{table}

We also define a conformance relation $\conformsTE{\Graph}{ts}{\typing}{te}$ which declares that the triples $ts$ in graph $\Graph$ conform to the triple expression $te$ with the shape assignment $\typing$ using the rules~\ref{sem:WShEx_TEs} which takes into account qualifier specifiers. 

\begin{table}[h!]
\begin{tabular}{c}
\inference[$EachOf$]
  { (ts_1,ts_2)\in\partition{ts} & 
    \conformsTE{\Graph}{ts_1}{\typing}{te_1} &
    \conformsTE{\Graph}{ts_2}{\typing}{te_2} 
  }
  {\conformsTE{\Graph}{ts}{\typing}{\eachOf{te_1}{te_2}}}

\\ \\

\inference[$OneOf_1$]
  { \conformsTE{\Graph}{ts}{\typing}{te_1} }
  {\conformsTE{\Graph}{ts}{\typing}{\oneOf{te_1}{te_2}} }

\hspace{1cm} 
\inference[$OneOf_2$]
  { \conformsTE{\Graph}{ts}{\typing}{te_2} }
  {\conformsTE{\Graph}{ts}{\typing}{\oneOf{te_1}{te_2}}}

\\ \\
\inference[$Star_1$]
  {}
  {\conformsTE{\Graph}{\emptyGraph}{\typing}{te*}}
\\ \\
\inference[$Star_2$]
  {(ts_1,ts_2)\in\partition{ts} & 
    \conformsTE{\Graph}{ts_1}{\typing}{te} &
    \conformsTE{\Graph}{ts_2}{\typing}{te*}}
  {\conformsTE{\Graph}{ts}{\typing}{te*}}

\\ \\

\inference[$TripleConstraint$]
  {ts = \{\quadruple{x}{p}{y}{s}\} & 
   \conforms{\Graph}{y}{\typing}{@\lbl} &
   \conformsQs{\Graph}{s}{\typing}{qs}
  }
  {\conformsTE{\Graph}{ts}{\typing}{\arc{p}{@\lbl}\,\,qs}}

\end{tabular}
\caption{Inference rules for WShEx triple expressions} \label{sem:WShEx_TEs}
\end{table}

Finally, the conformance relationship $\conformsQs{\Graph}{s}{\typing}{qs}$ 
 between a graph $\Graph$ 
 a set $s\in{}P\times{}V$ of property-value elements, 
 a shape assignment $\typing$ and 
 a qualifier specifier $qs$ is defined with the rules~\ref{sem:WShEx_Qs}. 

\begin{table}[h!]
\begin{tabular}{c}
\inference[$OpenQs$]
  {s'=\{(p,v)\in{}s|p\in{}preds(ps)\} &
   \conformsQs{\Graph}{s'}{\typing}{ps}
  }
  {\conformsQs{\Graph}{s}{\typing}{\openQs{ps}}} 

\hspace{0.4cm}

\inference[$CloseQs$]
  {\conformsQs{\Graph}{s}{\typing}{ps}}
  {\conformsQs{\Graph}{s}{\typing}{\closeQs{ps}}} 

\\ \\

\inference[$EachOfQs$]
  {\conformsQs{\Graph}{s}{\typing}{ps_1} &
   \conformsQs{\Graph}{s}{\typing}{ps_2}
  }
  {\conformsQs{\Graph}{s}{\typing}{\eachOfQs{ps_1}{ps_2}}} 
  
\\ \\

\inference[$OneOfQs_1$]
  {\conformsQs{\Graph}{s}{\typing}{ps_1}}
  {\conformsQs{\Graph}{s}{\typing}{\oneOfQs{ps_1}{ps_2}}} 
  
\hspace{0.5cm}

\inference[$OneOfQs_2$]
  {\conformsQs{\Graph}{s}{\typing}{ps_2} }
  {\conformsQs{\Graph}{s}{\typing}{\oneOfQs{ps_1}{ps_2}}} 
  
\\ \\
\hspace{-0.8cm}
\inference[$StarQs_1$]
  {}
  {\conformsQs{\Graph}{\emptyset}{\typing}{ps*}} 
  
\hspace{0.4cm}

\inference[$StarQs_2$]
  {(s_1,s_2)\in{}\partition{s} & 
   \conformsQs{\Graph}{s_1}{\typing}{ps} &
   \conformsQs{\Graph}{s_2}{\typing}{ps*}
  }
  {\conformsQs{\Graph}{s}{\typing}{ps*}} 
  
\\ \\

\inference[$EmptyQs$]
  {}
  {\conformsQs{\Graph}{\emptyset}{\typing}{\epsilon}} 
  
\hspace{0.5cm}

\inference[$PropertyQs$]
  { s = \{(p,v)\} &
    \conforms{\Graph}{v}{\typing}{@\lbl}
  }
  {\conformsQs{\Graph}{s}{\typing}{p:@\lbl}} 
  
\end{tabular}
\caption{Inference rules for WShEx qualifier expressions} \label{sem:WShEx_Qs}
\end{table}

%% file: 40_KGSubSets.tex
\section{Knowledge Graphs Subsets} \label{sec:KGSubSets}
In this section we review several approaches to create knowledge graphs subsets.
Although we will focus on Wikibase graphs subsets,
 the approaches described can also be applied to 
 RDF-based graphs and property graphs.
 
\subsection{Wikibase Subsets: Formal definition}

The following definition of Wikibase subset is based on the wikibase graphs definition
 given at section~\ref{sec:WikibaseDataModel_FormalDefinition}.

\begin{definition}[Wikibase subset] \label{def:wikibaseSubset}
Given a wikibase graph $\Graph=\langle\ItemSet,\PropSet,\DataValueSet,\StmtSet\rangle$,
 a wikibase subgraph is defined as $\Graph'=\langle\ItemSet',\PropSet',\DataValueSet',\StmtSet'\rangle$
 such that:
 $\ItemSet'\subseteq\ItemSet$,
 $\PropSet'\subseteq\PropSet$,
 $\DataValueSet'\subseteq\DataValueSet$ and 
 $\StmtSet'\subseteq\StmtSet$
\end{definition}

\begin{example}[Example of wikibase subgraph]
Given the wikibase graph from example~\ref{example:WikibaseGraph}  $\Graph'=\langle\ItemSet',\PropSet',\DataValueSet',\StmtSet'\rangle$ where 
\begin{equation*}
\begin{aligned}
\ItemSet' & =\{ \timBl, \London, \CERN \} \\
\PropSet' & =\{ \birthPlace, \employer, \start\} \\
\DataValueSet & = \{ 1980, 1984 \} \\
\StmtSet & = 
\{ (\timBl, \birthPlace, \London, \{\}), \\
 & (\timBl, \employer, \CERN, \{\start:1980 \}), \\
 & (\timBl, \employer, \CERN, \{\start:1984\}) \} \\
\end{aligned}
\end{equation*} 
is a wikibase subgraph of $\Graph$.
\end{example}

\subsection{Entity-generated subsets}

Wikibase subgraphs can be generated from a set of entities (items or properties), 
 where we collect the subgraph associated with those entities.

\begin{definition}[Item-generated subgraph] \label{def:RestrictedWikibaseSubset}
Given a wikibase graph $\Graph=\langle\ItemSet,\PropSet,\DataValueSet,\StmtSet\rangle$ and
 a subset of items $\ItemSet_s\subset{}\ItemSet$ 
 \index{item generated subgraph}
 generates an \emph{item-generated subgraph} $\langle\ItemSet',\PropSet',\DataValueSet',\StmtSet'\rangle$ 
 such that:
 \begin{flalign*}
 \ItemSet' &= \{q\in\ItemSet\mid(q,\_,\_,\_)\vee(\_,\_,q,\_)\in\StmtSet'\} \\
 &\cup \{ q\in\ItemSet\mid(\_,\_,\_,q_s)\in\StmtSet'\wedge(\_,q)\in{}q_s\} \\
 \PropSet' &= \{p\in\PropSet\mid(\_,p,\_,\_)\in\StmtSet'\} \\
 &\cup \{p\in\PropSet\mid(\_,\_,\_,q_s)\in\StmtSet'\wedge(p,\_)\in{}q_s\} \\ 
 \DataValueSet'&= \{d\in\DataValueSet\mid(\_,\_,d,\_)\in\StmtSet'\} \\
 &\cup \{d\in\DataValueSet\mid(\_,\_,\_,q_s)\in\StmtSet'\wedge(\_,d)\in{}q_s\} \\
\StmtSet' &= \{(q,\_,\_,\_)\in\StmtSet\mid{}q\in\ItemSet_s\} \\
 &\cup \{(\_,\_,q,\_)\in\StmtSet\mid{}q\in\ItemSet_s\} \\
 &\cup \{(\_,\_,\_,q_s)\in\StmtSet\wedge{}\exists{}q\in\ItemSet_s\mid(\_,q)\in{}q_s \} 
 \end{flalign*}
 
\end{definition}

Notice that the item-generated subgraph usually contains more items than the items provided by $\ItemSet_s$.

\begin{example}[Example of item-generated subgraph]

Given the wikibase graph from example~\ref{example:WikibaseGraph} and $\ItemSet_s=\{timBl\}$
 the item generated subgraph is:

\begin{flalign*}
\ItemSet' &= \{ \timBl, \CERN, \vintCerf, \PA \} &\\
\PropSet' &= \{ \birthDate, \birthPlace, \employer, \awarded, &\\ 
         &  \start, \pEnd, \togetherWith\} &\\
\DataValueSet' &= \{ 1984,1994,1980,1955\} &\\
\StmtSet' &= \{ (\timBl, \birthDate, 1955, \{\}), &\\ 
 & (\timBl, \birthPlace, \London, \{\}), &\\
 & (\timBl, \employer, \CERN, \{\start:1980, \pEnd:1980\}), &\\
 & (\timBl, \employer, \CERN, \{\start:1984, \pEnd:1994\}), &\\
 & (\timBl, \awarded, \PA, \{\togetherWith:\vintCerf\}), &\\
 & (\vintCerf, \awarded, \PA, \{\togetherWith: \timBl \}) \} &\\
\end{flalign*}
 
\end{example}

\begin{definition}[Property-generated subgraph] \label{def:PropertyGeneratedSubgraph}
Given a wikibase graph $\Graph=\langle\ItemSet,\PropSet,\DataValueSet,\StmtSet\rangle$,
with the set of entities $\EntitySet=\ItemSet\cup\PropSet$
 a subset of properties $\PropSet_s\subset{}\PropSet$ 
 \index{property generated subgraph}
 generates a \emph{property generated subgraph} $\langle\ItemSet',\PropSet',\DataValueSet',\StmtSet'\rangle$ 
 such that:
 \begin{flalign*}
 \ItemSet' &= \{q\in\ItemSet\mid\exists{}p\in\PropSet_s\mid(q,p,\_,\_)\in\StmtSet\} \\
  &\cup \{q\in\ItemSet\mid\exists{}p\in\PropSet_s\mid(\_,p,q,\_)\in\StmtSet\} \\
  &\cup \{q\in\ItemSet\mid(\_,\_,\_,q_s)\in\StmtSet\wedge\exists{}p\in\PropSet_s\mid(p,\_)\in{}q_s\} \\
 \PropSet' &= \{p\in\PropSet_s\mid(\_,p,\_,\_)\in\StmtSet\} \\ 
  &\cup \{p\in\PropSet_s\mid\exists{}q_s\mid(\_,\_,\_,qs)\in\StmtSet\wedge(p,\_)\in{}q_s\} \\ 
 \DataValueSet'&= \{d\in\DataValueSet\mid\exists{}p\in\PropSet_s\mid(\_,p,d,\_)\in\StmtSet\} \\
 &\cup \{d\in\DataValueSet\mid(\_,\_,\_,q_s)\in\StmtSet\wedge\exists{}p\in\PropSet_s\mid(p,d)\in{}q_s\} \\
\StmtSet' &= \{(\_,p,\_,\_)\in\StmtSet\mid{}p\in\PropSet_s\} \\
 &\cup \{(\_,\_,\_,q_s)\in\StmtSet\mid\exists{}p\in\PropSet_s\mid(p,\_)\in{}q_s\}\\ 
 \end{flalign*}
 
\end{definition}

The property generated subgraph usually contains more properties than the properties provided by $\PropSet_s$.

\begin{example}[Example of property-generated subgraph]
Given the wikibase graph from example~\ref{example:WikibaseGraph} and $\PropSet_s=\{birthDate, togetherWith\}$
 the property generated subgraph is:

\begin{flalign*}
\ItemSet' &= \{ \timBl, \vintCerf, \PA \} &\\
\PropSet' &= \{ \birthDate, \awarded, \togetherWith \} &\\ 
\DataValueSet' &= \{ 1955 \} &\\
\StmtSet' &= \{ (\timBl, \birthDate, 1955, \{\}), &\\ 
 & (\timBl, \awarded, \PA, \{\togetherWith:\vintCerf\}), &\\
 & (\vintCerf, \awarded, \PA, \{\togetherWith: \timBl \}) \} &\\
\end{flalign*}
 
\end{example}

\index{Datatype-generated subgraph}
Notice that it is possible to define a \emph{Datatype-generated subgraph} in a similar way than the previous definitions.

\begin{definition}[Entity-generated subgraph]
Given a subset of entities $\EntitySet_s\subset\ItemSet\cup\PropSet$, the entity-generated subgraph is defined as the union of the item-generated subgraph with all the items in $\EntitySet_s$ and
 the property-generated subgraph with all the properties in $\EntitySet_s$.
\end{definition}

\subsection{Simple Matching-generated subsets}

\begin{definition}[Matching expression]
Given a wikibase graph 
$\Graph=\langle\ItemSet,\PropSet,\DataValueSet,\StmtSet\rangle$ 
where $\EntitySet=\ItemSet\cup\PropSet$
and $\ValueSet=\EntitySet\cup\DataValueSet$, 
a matching expression $M_s$ is a set of matchers
where each matcher $m$ follows the grammar:

\begin{tabular}{ccll}
  $m$   & ::= &  subject($e$)  & Subject $e\in\EntitySet$\\
         & $|$ & property($p$) & Property $p\in\PropSet$ \\
         & $|$ & value($v$) & Value $v\in\ValueSet$\\
         & $|$ & qualifier($p$,$v$)  & Qualifier with property $p\in\PropSet$ and value $v\in\ValueSet$\\  
         & $|$ & qualifiedProp($p$)  & Qualifier with property $p\in\PropSet$ \\  
         & $|$ & qualifiedValue($v$)  & Qualifier with value $v\in\ValueSet$ \\  
\end{tabular}

\end{definition}

\begin{example}[Example of a matching expression]
An example of a matching expression is $M_s = \{ \text{property(country)}, \text{qualifiedProp(togetherWith)} \}$
\end{example}

\begin{definition}[Matching-generated subgraph]
Given a matching expression $M_s$ over a wikibase graph 
 $\Graph=\langle\ItemSet,\PropSet,\DataValueSet,\StmtSet\rangle$ 
  we can define the matching-generated subgraph as a wikibase graph 
  $\Graph'=\langle\ItemSet'\PropSet'\DataValueSet'\StmtSet'\rangle$ such that:

\begin{flalign*}
 \ItemSet' &= \{ q\in\ItemSet\mid (q,\_,\_,\_)\in\StmtSet'\cup\{q\in\ItemSet\mid(\_,\_,q,\_)\in\StmtSet'\}\\
  &\cup \{q\in\ItemSet\mid(\_,\_,\_,q_s)\in\StmtSet'\wedge(\_,q)\in{}q_s\} \\
 \PropSet' &= \{ p\in\PropSet\mid(\_,p,\_,\_)\in\StmtSet' 
 \cup\{p\in\PropSet\mid(\_,\_,\_,q_s)\in\StmtSet' \wedge(p,\_)\in{}q_s\} \\
 \DataValueSet' &= \{ d\in\DataValueSet\mid(\_,\_,d,\_)\in\StmtSet'\} 
  \cup\{d\in\DataValueSet\mid(\_,\_,\_,q_s)\in\StmtSet'\wedge(\_,d)\in{}q_s\}\\
 \StmtSet' &= \{ (q, \_, \_, \_)\in\StmtSet \mid \text{subject($q$)}\in{}M_s \} \\
   &\cup \{ (\_, p, \_, \_)\in\StmtSet \mid \text{property($p$)}\in{}M_s \}   \\
   &\cup \{ (\_, \_, v, \_)\in\StmtSet \mid \text{value($v$)}\in{}M_s \}   \\
   &\cup \{ (\_, \_, \_, q_s)\in\StmtSet \mid \text{qualifier($p$,$v$)}\in{}M_s \wedge\exists(p,v)\in{}q_s\}   
   &\cup \{ (\_, \_, \_, q_s)\in\StmtSet \mid \text{qualifiedProp($p$)}\in{}M_s \wedge\exists(p,\_)\in{}q_s\}   
   &\cup \{ (\_, \_, \_, q_s)\in\StmtSet \mid \text{qualifiedValue($v$)}\in{}M_s \wedge\exists(\_,v)\in{}q_s\}   
\end{flalign*}
  
\end{definition}

\begin{example}[Example of matching-generated subgraph]
Given the wikibase graph $\Graph$ of example~\ref{example:WikibaseGraph} and the matching-expression $M_s$ in example~\ref{ex:Matching-expression}, the matching-generated subgraph of $\Graph$ from $M_s$ is the wikibase graph $\Graph'=\langle\ItemSet',\PropSet',\DataValueSet',\StmtSet'\rangle$ such that:

\begin{flalign*}
\ItemSet' &= \{ \PA, \Spain, \London, \UK, \timBl, \vintCerf \} &\\
\PropSet' &= \{ \country, \awarded, \togetherWith \} &\\ 
\DataValueSet' &= \{ \} &\\
\StmtSet' &= \{ (\timBl, \awarded, \PA, \{ \togetherWith : \vintCerf \}), &\\ 
 & (\vintCerf, \awarded, \PA, \{\togetherWith: \timBl \}) &\\
 & (\PA, \country, \Spain, \{ \}) &\\
 & (\London, \country, \UK, \{ \}) \} &\\
\end{flalign*}

\end{example}

The matching approach is followed by 
 WDumper~\footnote{\url{https://github.com/bennofs/wdumper}} and WDSub\footnote{\url{https://github.com/weso/wdsub}}. 
 
 WDumper defines the expected patterns using a JSON configuration file that describes them or 
  filling a web form which internally generates the JSON file.
  
In the case of WDSub, the input format is a WShEx file with a set of shapes and the system processes a Wikidata dump trying to match each entity 
  with any of the Shapes defined in the WShEx file. 
 The algorithm employed in WDSub to generate a matching expression from a Shape Expression is the following:
 

\subsection{ShEx-based Matching generated subsets}  \label{sec:ShExBasedMatchingSubsets}

ShEx-based matching consists on taking as input a WShEx schema $\schema$ and 
 include in the generated subset the nodes whose neighborhood matches any of the shapes from $\schema$ after replacing 
 any shape references by a condition that always returns true. 
The goal of this approach is to use ShEx as a basic description language of the topology of nodes ignoring shape references so the algorithm can be used to check dumps that contain include the information about a node and its neighborhood in a single line. 
In this way, the subset generator only needs to traverse the dump sequentially one time.

\begin{example}[ShEx-based matching]
Giveb the following WShEx Schema:

\begin{tabular}{cm{0.2em}m{0.2em}l}
\LabelSet & = & \{ & $Researcher$, $Place$, $Country$, $Date$, $Human$\} \\
\schemaDef($Researcher$) & = & \{ & \arc{\instanceOf}{@$Human$}; \\
                     &   &  & \arc{\birthDate}{@$Date$}?; \\
                     &   &  & \arc{\birthPlace}{@$Place$} \\
                      & & \} & \\
\schemaDef($Place$)  & = & \{ & \arc{\country}{@$Country$}\} \\
\schemaDef($Date$)  & = &  & $\in{}xsd:date$\\
\schemaDef($Human$)  & = &  & $\in{}\{\Human\}$\\
\end{tabular}

The result of ShEx-based matching on example~\ref{example:WikibaseGraph} is:

\begin{tabular}{ccl}
$\StmtSet$ &= \{ & (\timBl, \instanceOf, \Human, \{\}), \\ 
 & & (\timBl, \birthDate, 1955, \{\}), \\ 
 & & (\timBl, \birthPlace, \London, \{\}), \\
 & & (\London, \country, \UK, \{\}), \\
 & & (\vintCerf, \instanceOf, \Human, \{\}) \\
 & & (\vintCerf, \birthPlace, \newHaven, \{\}) \\
 & \} & \\
\end{tabular}

Notice that \vintCerf is included although the node doesn't conform to the shape person because it has a \birthPlace declaration whose value is \newHaven but there is no \country property for \newHaven.

In the previous example, the ShEx-based matching consisted on validating each node with any of the following shapes:

\begin{tabular}{cm{0.2em}m{0.2em}l}
\schemaDef($Person$) & = & \{ & \arc{\instanceOf}{\lstinline|true|}; \\
                     &   &  & \arc{\birthDate}{\lstinline|true|}?; \\
                     &   &  & \arc{\birthPlace}{\lstinline|true|} \\
                      & & \} & \\
\schemaDef($Place$)  & = & \{ & \arc{\country}{\lstinline|true|}\} \\
\end{tabular}

Notice that if the original ShEx schema had included the following shape:

\begin{tabular}{cm{0.2em}m{0.2em}l}
\schemaDef($Country$) & = & \{ & \} \\
\end{tabular}

Then, every node would be included in the generated subset because every node would match the $Country$ shape. 

\end{example} 

ShEx-based matching generation has been implemented in WDSub\footnote{\url{https://github.com/weso/wdsub}}.

\subsection{ShEx + Slurp generated subsets}

The concept of \emph{slurp} was introduced in the shex.js\footnote{\url{https://github.com/shexjs/shex.js}} implementation as a mechanism to 
 collect the nodes and triples visited during validation. 
 
 In this way, if we collect that data, the result will be a subset of the graph which 
 contains the portion of the graph that relates to a given ShEx schema.
 Although the slurp option was not formally defined, we can define it modifying the semantics of ShEx adding a new parameter to the conformance relationship. 
 
 We define a conformance relation parameterized by a shape assignment $\slurpSE{\Graph}{n}{\typing}{se}{\Graph'}$ 
  with the meaning that node $n$ in graph $\Graph$ conforms to shape expression $se$ with shape assignment $\typing$ 
  and generates a slurp graph $\Graph'$. The conformance relation follows the rules~\ref{slurp:WShEx_SE}.
 
 \begin{table}[h!]
 \begin{tabular}{c}
 \inference[$Cond$]
   {cond(n)=true} 
   {\slurpSE{\Graph}{n}{\typing}{cond}{\langle\{n\},\{\}\rangle}}
 
   \hspace{1cm}
 \inference[$AND$]
   {\slurpSE{\Graph}{n}{\typing}{se_1}{\Graph_1} & 
    \slurpSE{\Graph}{n}{\typing}{se_2}{\Graph_2}
   } 
   {\slurpSE{\Graph}{n}{\typing}{se_1 \text{ \texttt{AND} } se_2}{\Graph_1\cup\Graph_2}}
 \\ \\
   
 %
 %
 
   \inference[$ClosedShape$]
   { neighs(n,\Graph) = ts & 
    \slurpTE{\Graph}{ts}{\typing}{s'}{\Graph'}
   } 
   {\slurpSE{\Graph}{n}{\typing}{$\c|CLOSED|\,$s'}{\Graph'}}  
   \\ \\
 
   \inference[$OpenShape$]
   { ts = \{\triple{x}{p}{y} \in neighs(n,\Graph) \mid p \in \props{te}\} &
     \slurpTE{\Graph}{ts}{\typing}{s'}{\Graph'}
   } 
   {\slurpSE{\Graph}{n}{\typing}{s'}{\Graph'}}
   \\ \\
 \end{tabular}
 \caption{Inference rules for WShEx+slurp shape expressions} \label{slurp:WShEx_SE}
 \end{table}
 
 We also define a conformance relation $\slurpTE{\Graph}{ts}{\typing}{te}{\Graph'}$ 
  which declares that the triples $ts$ in graph $\Graph$ conform to the triple expression $te$ with the shape assignment $\typing$ 
  generating a slurp $\Graph'$. The relation is defined using the rules~\ref{slurp:WShEx_TEs}.

 \begin{table}[h!]
 \begin{tabular}{c}
 \inference[$EachOf$]
   { (ts_1,ts_2)\in\partition{ts} & 
     \slurpTE{\Graph}{ts_1}{\typing}{te_1}{\Graph_1} &
     \slurpTE{\Graph}{ts_2}{\typing}{te_2}{\Graph_2} 
   }
   {\slurpTE{\Graph}{ts}{\typing}{\eachOf{te_1}{te_2}}{\Graph_1\cup\Graph_2}}
 
 \\ \\
 
 \inference[$OneOf_1$]
   { \slurpTE{\Graph}{ts}{\typing}{te_1}{\Graph_1} }
   { \slurpTE{\Graph}{ts}{\typing}{\oneOf{te_1}{te_2}}{\Graph_1} }
 
 \hspace{1cm} 
 \inference[$OneOf_2$]
   { \slurpTE{\Graph}{ts}{\typing}{te_2}{\Graph_2} }
   { \slurpTE{\Graph}{ts}{\typing}{\oneOf{te_1}{te_2}}{\Graph_2} }
 
 \\ \\
 \inference[$Star_1$]
   {}
   {\slurpTE{\Graph}{\emptyGraph}{\typing}{te*}{\emptyGraph}}
 \\ \\
 \inference[$Star_2$]
   {(ts_1,ts_2)\in\partition{ts} & 
     \slurpTE{\Graph}{ts_1}{\typing}{te}{\Graph_1} &
     \slurpTE{\Graph}{ts_2}{\typing}{te*}{\Graph_2}
   }
   {\slurpTE{\Graph}{ts}{\typing}{te*}{\Graph_1\cup\Graph_2}}
 
 \\ \\
 
 \inference[$TripleConstraint$]
   {ts = \{\quadruple{x}{p}{y}{s}\} & 
    \slurpSE{\Graph}{y}{\typing}{@\lbl}{\langle\VertSet,\EdgeSet\rangle} &
    \slurpQs{\Graph}{s}{\typing}{qs}{(qs',\Graph_{qs})}
   }
   {\slurpTE{\Graph}{ts}{\typing}{\arc{p}{@\lbl}\,\,qs}{\langle\VertSet\cup\{x\}\cup\{y\},\EdgeSet\cup(x,p,y,qs')\rangle\cup\Graph_{qs}}}
 
 \end{tabular}
 \caption{Inference rules for WShEx+slurp triple expressions} \label{slurp:WShEx_TEs}
 \end{table}
 
The conformance relationship $\slurpQs{\Graph}{s}{\typing}{qs}{(qs',\Graph')}$ 
   between a graph $\Graph$ 
   a set $s\in{}P\times{}V$ of property-value elements, 
   a shape assignment $\typing$ and 
   a qualifier specifier $qs$ generates a slurp that consists of a pair $(qs',\Graph')$ where $qs'$ is a set of qualifiers slurped
   and $\Graph'$ is the graph slurped. It is defined according to the rules~\ref{slurp:WShEx_Qs}.

 \begin{table}
 \begin{tabular}{c}
 \inference[$OpenQs$]
   {s'=\{(p,v)\in{}s|p\in{}preds(ps)\} &
    \slurpQs{\Graph}{s'}{\typing}{ps}{(qs,\Graph')}
   }
   {
    \slurpQs{\Graph}{s}{\typing}{\openQs{ps}}{(qs,\Graph')}
   } 
 \\ \\
 
 \inference[$CloseQs$]
   {
    \slurpQs{\Graph}{s}{\typing}{ps}{(qs,\Graph')} 
   }
   {
    \slurpQs{\Graph}{s}{\typing}{\closeQs{ps}}{(qs,\Graph')}
   } 
 
 \\ \\
 
 \inference[$EachOfQs$]
   {
    \slurpQs{\Graph}{s}{\typing}{ps_1}{(qs_1,\Graph_1)} &
    \slurpQs{\Graph}{s}{\typing}{ps_2}{(qs_2,\Graph_2)}
   }
   {
    \slurpQs{\Graph}{s}{\typing}{\eachOfQs{ps_1}{ps_2}}{(qs_1\cup\qs_2,\Graph_1\cup\Graph_2)}
   } 
   
 \\ \\
 
 \inference[$OneOfQs_1$]
   {
    \slurpQs{\Graph}{s}{\typing}{ps_1}{(qs_1,\Graph_1)}
   }
   {
    \slurpQs{\Graph}{s}{\typing}{\oneOfQs{ps_1}{ps_2}}{(qs_1,\Graph_1)}
   } 
   
 \hspace{0.5cm}
 
 \inference[$OneOfQs_2$]
   {
    \slurpQs{\Graph}{s}{\typing}{ps_2}{(qs_2,\Graph_2)} }
   {
    \slurpQs{\Graph}{s}{\typing}{\oneOfQs{ps_1}{ps_2}}{(qs_2,\Graph_2)}
   } 
   
 \\ \\

 \inference[$StarQs_1$]
   {}
   {
    \slurpQs{\Graph}{\emptyset}{\typing}{ps*}{(\{\},\emptyGraph)}
   } 

 \\  \\
 
 \inference[$StarQs_2$]
   {(s_1,s_2)\in{}\partition{s} & 
    \slurpQs{\Graph}{s_1}{\typing}{ps}{(qs_1,\Graph_1)} &
    \slurpQs{\Graph}{s_2}{\typing}{ps*}{qs_2,\Graph_2}
   }
   {
    \slurpQs{\Graph}{s}{\typing}{ps*}{(qs_1\cup{}qs_2,\Graph_1\cup\Graph_2)}
   } 
   
 \\ \\
  
 \inference[$EmptyQs$]
   {}
   {
    \slurpQs{\Graph}{\emptyset}{\typing}{\epsilon}{(\{\},\emptyGraph)}
   } 
   
 \hspace{0.5cm}
 
 \inference[$PropertyQs$]
   { s = \{(p,v)\} &
     \slurpSE{\Graph}{v}{\typing}{@\lbl}{\Graph'}
   }
   {
    \slurpQs{\Graph}{s}{\typing}{p:@\lbl}{(\{p:v\},\Graph')}
   } 
   
 \end{tabular}
 \caption{Inference rules for WShEx+slurp qualifiers} \label{slurp:WShEx_Qs}
 \end{table}
 
\begin{example}[ShEx+Slurp example] \label{example:ShExSlurp}

Given the following WShEx Schema:

\begin{tabular}{cm{0.2em}m{0.2em}l}
\LabelSet & = & \{ & $Researcher$, $Place$, $Country$, $Date$, $Human$\} \\
\schemaDef($Researcher$) & = & \{ & \arc{\instanceOf}{@$Human$}; \\
                     &   &  & \arc{\birthDate}{@$Date$}; \\
                     &   &  & \arc{\birthPlace}{@$Place$} \\
                      & & \} & \\
\schemaDef($Place$)  & = & \{ & \arc{\country}{@$Country$}\} \\
\schemaDef($Country$)  & = & \{ & \} \\
\schemaDef($Date$)  & = &  & $\in{}xsd:date$\\
\schemaDef($Human$)  & = &  & $\in{}\{\Human\}$\\
\end{tabular}

The result of running the ShEx+Slurp on example~\ref{example:WikibaseGraph} is:

\begin{tabular}{ccl}
$\rho$ &= \{ & (\timBl, \instanceOf, \Human, \{\}), \\ 
 & & (\timBl, \birthDate, 1955, \{\}), \\ 
 & & (\timBl, \birthPlace, \London, \{\}), \\
 & & (\London, \country, \UK, \{\}), \\
\end{tabular}

The main difference between this approach and the previous one is that it retrieves the valid subset according to the ShEx schema. In this case, the node \vintCerf is not generated because the value of the property \birthPlace is \newHaven and it has no \country declaration, so \newHaven doesn't conform to the $Place$ shape and subsequently, \vintCerf doesn't conform to the $Person$ shape as they are declared in that Schema.

\end{example} 
 
Although the ShEx+Slurp approach has not yet been implemented for WShEx, is has already been implemented for ShEx in shex.js and in PyShEx~\footnote{\url{https://github.com/hsolbrig/PyShEx}}.
 
One problem of this approach is that it is difficult to scale as it needs to traverse the graph while validating and collecting the slurped graph. 
 The complexity also increases if the implementation wants to adjust the collected triples when checking the different partitions of a node neighborhood. If one of the partitions fails, following the definition it would need to discard the corresponding portion of the graph, which would make the whole process more complex. 
 In practice, implementations just collect the visited nodes and triples without discarding the ones that shouldn't be part of the result.

\subsection{ShEx + Pregel generated subsets}

Pregel~\cite{Malewicz2010} has been proposed as an scalable computational model created by Google to 
 handle large graphs. 
 It is based on Bulk Synchronous Parallel (BSP) model which simplifies parallel programming having
  different computation and communication phases. 
 Pregel is an iterative algorithm where each phase is called a superstep. 
 Following the lemma \emph{think like a vertex}, 
  it is a vertex-centric abstraction where at each superstep, 
  a vertex executes a user defined function (called vertex program) which can update its status and later sends messages 
  to neighbors along graph edges. 
  Supersteps end with a synchronization barrier that guarantees that messages sent at one superstep 
   are received at the beginning of the next superstep.
  Vertices may change status between active and inactive and the algorithm terminates when all vertices are inactive and no more messages are sent. 
  
GraphX was proposed in 2014 as a graph processing framework embedded in Apache Spark. Its API includes a variant of 
 Pregel which is used to implement several graph algorithms like PageRank, connected components, triangle counting, etc. 
 
GraphX defines an API for graphs based on RDDs (resilient distributed datasets). 
An $RDD[\ValueSet]$ is an abstraction of a collection of values of type $\ValueSet$ 
which are immutable and can be partitioned to run data-parallel operations like \emph{map} and \emph{reduce}.

A graph $Graph[\VertSet,\EdgeSet]$ represents and abstraction of vertices with values of type $\ValueSet$ 
and edges of type $\EdgeSet$ where internally the vertices are represented as 
$RDD[(Id,\ValueSet)]$, i.e. a collection of a tuple with an $Id$ (a $Long$ value) and a $\ValueSet$, 
and edges are represented as $RDD[(Id,Id,\EdgeSet)]$, i.e. a triple where the first and second components are the $Id$ of the source and destiny respectively, 
 and the third component is the edge property $p\in\EdgeSet$.
A graph $Graph[\VertSet,\EdgeSet]$ also provides what is called a \emph{triplets} view which represents edges as collections of triplets of the form $RDD[(\VertSet, \EdgeSet, \VertSet)]$. A triplet $t$ will be denoted by the type \c|Triplet| and provides access to the source vertex (using \c|t.srcAttr|), the destiny (\c|t.dstAttr|) and the edge property (\c|t.attr|).

GraphX provides several built-in operators for graphs\footnote{\url{https://spark.apache.org/docs/latest/graphx-programming-guide.html}}. 
 We will use the following in the rest of the paper:
 
\begin{itemize}
\item \c|mapVertices|(\c|g|: \c|Graph|[$\VertSet$,$\EdgeSet$], \c|f|: (\c|Id|,$\ValueSet$)$\rightarrow\ValueSet$): \c|Graph|[$\ValueSet$,$\EdgeSet$] maps every pair \c|(id,v)| in the vertices of \c|g| to \c|(id, f(v))|.
\item \c|mapReduceTriples|(\c|g|:\c|Graph|[\VertSet,\EdgeSet], 
 \c|m|: (\VertSet,\EdgeSet,\VertSet)$\rightarrow$[(\c|Id|,\MsgSet)], \c|r|:(\MsgSet,\MsgSet)$\rightarrow$\MsgSet):\c|RDD|[(Id,\MsgSet)], encodes the two-stage parallel computation process commonly known as mapReduce using the triplets view.
 It takes as parameters, a grapg \c|g|, 
 a map function \c|m| and a reduce function \c|r|. 
 
 In the first stage it applies the \c|m| to each triplet in the graph 
  to generate a list of messages that will be sent to the vertices identified a given \c|id|.
 
 In the second stage, it groups all the messages sent to a given vertex applying the reduce function \c|r| to each pair of messages.
 
\item \c|joinVertices|(\c|g|:\c|Graph|[\VertSet,\EdgeSet], \c|msgs|:\c|RDD|[(\c|Id|, \MsgSet)], \c|f|:(\c|Id|, \VertSet,\MsgSet)$\rightarrow$\VertSet): \c|Graph|[\VertSet,\EdgeSet], joins the collection of messages sent to a the vertices which have a value \c|(id,m)| with the vertex \c|v| identified by \c|id| and replaces that vertex by \c|f(id,v,m)|.
\end{itemize}

%
%

\index{superstep}
\index{Pregel}
The GraphX Pregel algorithm is defined iteratively where each iteration is usually called a superstep as follows:

\begin{algorithm}[!hbt]
\DontPrintSemicolon
\SetAlgoVlined
\SetArgSty{textnormal}
\SetKwInput{KwIn}{Input parameters}
\KwIn{
 \Block{\c|g|: \c|Graph|[\VertSet,\EdgeSet] \;
 \c|initialMsg|: \MsgSet \;
 \c|vProg|: (\c|Id|,\VertSet,\MsgSet)$\rightarrow$\VertSet \;
 \c|sendMsg|: \c|Triplet|$\rightarrow$[(\c|Id|,\MsgSet)] \;
 \c|mergeMsg|: (\MsgSet,\MsgSet)$\rightarrow$\MsgSet}
}
\KwOut{\c|g|:\c|Graph|[\VertSet,\EdgeSet]
}
\blockskip
\c|g| = \c|mapVertices|(\c|g|,$\lambda$(\c|id|,\c|v|)$\rightarrow$\c|vProg|(\c|id|,\c|v|,\c|initialMsg|))  \;
\c|msgs| = \c|mapReduceTriples|(\c|g|,\c|sendMsg|,\c|mergeMsg|) \;
\While{\c|size|(\c|msgs|)$>0$}{
 \c|g| = \c|joinVertices|(\c|g|,\c|msgs|,\c|vProg|) \;
 \c|msgs| = \c|mapReduceTriples|(\c|g|,\c|sendMsg|,\c|mergeMsg|)
}
\Return\c|g|
\caption{Pregel algorithm pseudocode as implemented in GraphX}
\end{algorithm}

It takes as input a \c|Graph|[\VertSet,\EdgeSet] and the following parameters:

\begin{itemize}
\item  \c|initialMsg|: initial message sent to all the vertices
\item \c|vprog| is the vertex program. 
 It is run by each vertex at the beginning of the algorithm using the \c|initialMsg| and in each superstep using the collected messages sent by the neighbors in the previous superstep.
\item \c|sendMsg| takes as parameter an triplet and returns an iterator with a pair \c|(id, msg)| where \c|id| represents the id of the vertex 
 which will receive the message and \c|msg| represents the message that will be sent.
\item \c|mergeMsg| is a function that defines how to merge 2 messages into one. 
 This function must be associative and commutative, and will be invoked to collect all the messages that are sent to a vertex in each superstep. 
\end{itemize} 

We have implemented a ShEx validation algorithm based on the Pregel algorithm.
 The algorithm assumes that there is a ShEx schema $\langle\LabelSet,\schemaDef\rangle$ where each label $\lbl\in\LabelSet$ identifies a shape expression. 
 
The algorithm annotates each node $n\in{}\VertSet$ with a status map that represents the validation status with regards to some labels.
The new nodes in the graph will be tuples 
$(n,m)$ where 
 $n\in\VertSet$, and 
 $m:\LabelSet\mapsto{}Status$ associates a status for each shape label. 
 
A $Status$ is defined as: 

\begin{tabular}{ccll}
$Status$ & ::= & \Undefined    & Default status \\
    & $\mid$ & \Ok  & Node conforms\\
    & $\mid$ & \Failed  & Node doesn't conform \\
    & $\mid$ & \Pending & Requested to conform \\
    & $\mid$ & \WaitingFor{ds}{oks}{fs} & Waiting for some neighbours \\
    & & & $ds$ = list dependants neighbours \\
    & & & $oks$ = list of conformant neighbours \\
    & & & $fs$ = list of non conformant neighbours \\
    & & & where $ds, oks, failed\in\VertSet\times\PropSet\times\LabelSet$
\end{tabular}

\noindent{}The status can be 
$\Undefined$ if there is no information yet (this is the default value)
$\Ok$ if the node conforms to the shape identified by $\lbl$, 
$\Failed$ if it doesn't conform to the shape, 
$\Pending$ if the node has been requested to be validated with that label or
$\WaitingFor{ds}{oks}{failed}$ if the validation of node $n$ depends on
 the validation of a set of neighbour nodes $ds$. 
 Each neighbour node is represented by a triple $(v,p,l)$ where 
 $v$ is the neighbour node,
 $p$ is the property which links $n$ with $v$,
 and $l$ is the shape label that the node must conform. 
 During the validation, we may receive information that some of those neighbour nodes have been validated or failed. 
 That information is collected in the set
 $oks$ which is the set of conforming neighbour nodes and
 $failed$ is the set of failed neighbour nodes. 

A message can be represented as a map which assigns to each label the following requests: 

\begin{tabular}{ccll}
Msg & ::= & \Validate   &  Request to validate \\
    & $\mid$ & \Checked{oks}{fs} & Some neighbours have been checked \\
    & & & $oks$ = neighbours that have been checked as conformant \\
    & & & $fs$ = neighbours that have been checked as non-conformant \\
    & & & where $oks,fs\in\VertSet\times\PropSet\times\LabelSet$ \\
    & $\mid$ & \WaitFor{ds} & Request to wait for some neighbours \\
    & & & where $ds\in\VertSet\times\PropSet\times\LabelSet$
\end{tabular}

The ShEx+Pregel validation traversal is defined with the following pseudo-code.

\LinesNumberedHidden
\begin{algorithm}[!hbt]
\DontPrintSemicolon
\SetAlgoVlined
\SetArgSty{textnormal}
\SetKwInput{KwIn}{Input parameters}
\KwIn{
 \Block{\c|g|: \c|Graph|[\VertSet, \EdgeSet] \;
 \c|initialLabel|: \LabelSet \;
 \c|checkLocal|: (\LabelSet, \VertSet) $\rightarrow$ \Ok $\mid$ \Failed $\mid$ \Pending(\c|Set|[\LabelSet]) \;
 \c|checkNeighs|: (\LabelSet, \c|Bag|[(\EdgeSet, \LabelSet)],  \c|Set|[(\EdgeSet, \LabelSet)]) $\rightarrow$ \Ok$\mid$\Failed \;
 \c|tripleConstraints|: \LabelSet $\rightarrow$ \c|Set|[(\EdgeSet, \LabelSet)]}
}
\KwOut{\c|g|:\c|Graph|[(\VertSet, \LabelSet $\mapsto$ Status), \EdgeSet]
}
\blockskip
\c|gs| = \c|mapVertices|(\c|g|, $\lambda$(\c|id|, \c|v|)$\rightarrow$(id, (v, $\lambda$v$\rightarrow$\Undefined)))  \;
\c|gs| = \c|pregel|(\Validate, \c|gs|, \vProg, \c|sendMsg|, \c|mergeMsg|) \;
\c|gs| = \c|mapVertices|(\c|gs|, \c|checkUnsolved|) \;
\Return\c|gs| \;
\blockskip
\DefInline{\c|checkUnsolved|(v,m)}{
(v,m') where \; 
\ \ \ m'(\lbl) = $\begin{cases}
  \text{\c|checkNeighs|(\lbl, $\emptyset$, $\emptyset$)} & \text{if}\ m(\lbl)=\Pending \\ 
  \text{\c|checkNeighs|(\lbl, oks, fs $\cup{}$ ds)} & \text{if}\ m(\lbl)=\WaitingFor{ds}{oks}{fs}\} \\
  m(\lbl) & \text{otherwise} 
  \end{cases}$
}\\
\DefInline{\c|vProg|:(\c|Id|,\VertSet,\MsgSet)$\rightarrow$\VertSet}{...see~\ref{vProg_ShEx}}
\caption{Pregel-based ShEx validation pseudocode} \label{PSchema}
\end{algorithm}

The algorithm takes as input the parameters:

\begin{itemize}
\item \c|initialLabel| is the initial shape label that is requested to validate every node in the graph. 
 In Shape Expressions, this label is usually annotated with the \c|start| keyword.
\item \c|checkLocal| checks if the shape expression associated with a 
	label can validate a node locally. 
	It returns \Ok if the node validates without further dependencies, 
	\Failed, if it doesn't validate, and 
	\Pending($ls$) if the validation of the node depends on a list of shape labels $ls$. 
	
\item \c|checkNeighs| checks if the bag of neighbors of a node matches the
	 regular bag expression associated with the label in the schema.
\item \c|tripleConstraints| returns the list of triple constraints associated with
	 the shape expression indicated by the label.
\end{itemize} 

\index{Pregel}
The algorithm starts by mapping every node to the status which associates any label $\lbl\in\LabelSet$ to undefined ($\Undefined$).
After that, it runs the iterative Pregel algorithm using the \c|vProg|, \c|sendMsg| and \c|mergeMsg| functions defined as above. 
Once the Pregel algorithm finishes, it replaces the status of any node that is pending or waiting for some neighbours by a last check based on the current information of the neighbours, assuming that if the node didn't receive information that a pending neighbour has validated, it means that there was no evidence of it's validation, and it failed.

\index{vProg}
\c|vProg| changes the status map of a node with regards to a label when it receives a message for that label. It can be defined as:

\c|vProg|($id$,($n$,$m$), $msg$) = ($n$,$m'$)\ where $m'(\lbl)=m(\lbl)$ except for the cases indicated by the following rules:

\begin{table}
\begin{tabular}{c}
\inference[$$]
  {\fracEmpty{
    \msgSent{(n,m)}{\lbl}{\Validate}
    }{
    \status{m}{\lbl}=s\in\{\Undefined, \Pending\}
    } & \checkLocal{\lbl}{n}=r\in\{\Ok,\Failed\}
  } 
  {\status{m'}{\lbl}=r}
  
  \\ \\

\inference[$$]
  {\fracEmpty{\msgSent{(n,m)}{\lbl}{\Validate}}   
   {\status{m}{\lbl}=r\in\{\Undefined,\Pending\}
   } & \checkLocal{\lbl}{n}=\PendingLs
  } 
  {\fracEmpty{\status{m'}{l}=\Undefined}{\status{m'}{l'}=\Pending\,\,\forall{l'}\in{}ls}}

\\ \\  

\inference[$$]
  {\fracEmpty{
    \msgSent{(n,m)}{\lbl}{\Validate}}    
   {
    \status{m}{\lbl}=r\in\{\Ok,\Failed\}
   }
  } 
  {\status{m'}{\lbl}=r}

\\ \\  

\inference[$$]
  {\fracEmpty{
    \msgSent{(n,m)}{\lbl}{\Validate}
    }{
    \status{m}{\lbl}=\WaitingFor{ds}{oks}{fs}
  }
  } 
  {\status{m'}{\lbl}=\Ok}

\\ \\  

\inference[$$]
  {\fracEmpty{
   \msgSent{(n,m)}{\lbl}{\Checked{oks}{fs}}}
   { 
   \status{m}{\lbl}=\WaitingFor{ds}{oks'}{fs'}
   } & ds\setminus(oks\cup{}fs)\neq{}\emptyset
  } 
  {\status{m'}{\lbl}=\WaitingFor{ds}{oks\cup{}oks'}{fs\cup{}fs'}}

\\ \\  

\inference[$$]
  {\fracEmpty{\msgSent{(n,m)}{\lbl}{\Checked{oks}{fs}}}
   {\status{m}{\lbl}=\WaitingFor{ds}{oks'}{fs'}} & 
   ds\setminus(oks\cup{}fs)=\emptyset 
  } 
  {\status{m'}{\lbl}=\checkNeighs{\lbl}{oks\cup{}oks'}{fs\cup{}fs'}}

\end{tabular}\\
\label{vProg_ShEx}
\caption{Definition of \c|vProg| for Pregel-based ShEx validation}
\end{table} 

\begin{figure}
\centering
\includegraphics[width=0.8\textwidth]{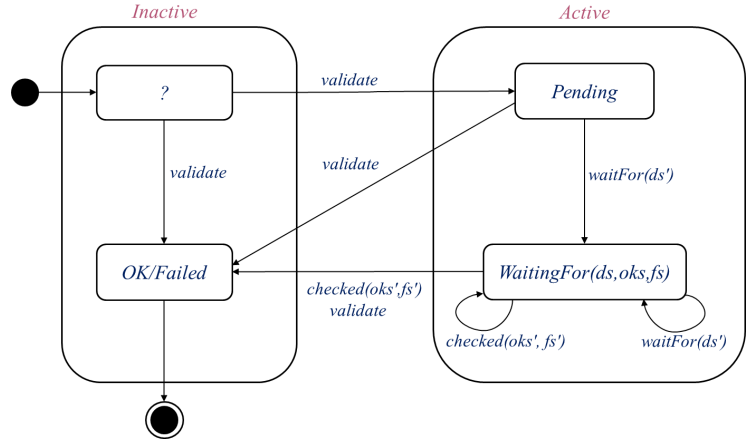}
\caption{State diagram representing the different states in vProg} \label{figure:StateDiagramVProg}
\end{figure}

Figure~\ref{figure:StateDiagramVProg} represents a state diagram which shows the different status that a node can have with regards to a shape label. 
 Initially, all nodes have status \Undefined{} until they get a message request to validate against some label. 
 If it is possible to validate locally those nodes, then they will go directly to the end state which can be \Ok{} or \Failed{}. Otherwise, if their validation depends on the neighbours, they will enter the status \Pending{} whose nodes are active in the Pregel algorithm and will be activated in the messages generation phase. 
 If they receive a request to wait for some other nodes to be validated, they will go to the state $\WaitingFor{ds}{oks}{fs}$ 
 which means that they are waiting for the status of the neighbour nodes $ds$. 
 
In subsequent phases, they can receive notifications that some of those neighbour nodes 
  have either been validated or not updating the corresponding values of $oks$ and $fs$. 
 Once all the pending neighbours have either been validated or failed, it will invoke $\checkNeighs{\lbl}{oks}{fs}$ to check if the regular expression matches taking into account which neighbours conform or don't conform and passing to the state $\Ok$ or $\Failed$ which is inactive.
 
 Once executed the Pregel algorithm, it is possible that some nodes are in state \Pending and don't receive any message, which means that their validation depends on the existence of some arcs pointing to some neighbours and they didn't receive messages from those arcs, i.e. there are no arcs in the graph. In that case, a last step in the algorithm checks if those nodes can validate with an empty neighbourhood. 

\index{sendMsg}
In order to define \c|sendMsg|($Triplet$):[($Id$,$Msg$)]\ we will use the notation 
 $\msgSent{(x,m_x)}{\lbl}{Msg}$ to represent that message $Msg$ is sent to the node $x$ with status map $m_x$ for label $\lbl$.

Table~\ref{table:sendMsg_ShEx} represents the rules that declare which messages are sent for each triplet view 
 which is represented as $\triple{(s,m_s)}{p}{(o,m_o)}$ where $(s,m_s)$ is the subject, $p$ the predicate and $(o,m_o)$ the object:

\begin{table}
\begin{tabular}{c}
\inference[$$]
  {
   \triple{(s,m_s)}{p}{(o,m_o)}\in\Graph & \status{m_s}{\lbl}=\Pending & \tcs{\lbl}{\schema}=\arc{p}{@\lbl'}
  } 
  {
    \fracEmpty{
     \msgSent{(s,m_s)}{\lbl}{\WaitFor{(o,p,\lbl')}}
    }{
     \msgSent{(o,m_o)}{\lbl}{\Validate}
    }
  }
  
  \\ \\

\inference[$$]
  {
   \triple{(s,m_s)}{p}{(o,m_o)}\in\Graph & \status{m_s}{\lbl}=\WaitingFor{ds}{oks}{fs} &
   (o,p,\lbl')\in{}ds &
   \status{m_o}{\lbl'}=\Ok
  } 
  {
     \msgSent{(s,m_s)}{\lbl}{\Checked{(o,p,\lbl')}{\emptyset}}
  }

\\ \\

\inference[$$]
  {
   \triple{(s,m_s)}{p}{(o,m_o)}\in\Graph & \status{m_s}{\lbl}=\WaitingFor{ds}{oks}{fs} &
   (o,p,\lbl')\in{}ds &
   \status{m_o}{\lbl'}=\Failed
  } 
  {
     \msgSent{(s,m_s)}{\lbl}{\Checked{\emptyset}{(o,p,\lbl')}}
  }

\end{tabular}\\
\label{table:sendMsg_ShEx}
\caption{Definition of \c|sendMsg| for Pregel-based ShEx validation}
\end{table} 

\index{mergeMsg}
Finally \c|mergeMsg| merges the messages that arrive to the same node and can be defined as:

\begin{tabular}{ccll}
\c|mergeMsg|(\msgSent{(n,m)}{\lbl}{msg_1}, \msgSent{(n,m)}{\lbl}{msg_2}) & = & \msgSent{(n,m)}{\lbl}{msg_1\oplus{}msg_2} \\
\end{tabular}

where

\begin{tabular}{rll}
 $\Validate \oplus y$ & = $y$ & \\
 $\Validate \oplus \Checked{oks}{fs}$ & = $\Checked{oks}{fs}$ & \\
 $\Validate\oplus \WaitFor{ds}$ & = $\WaitFor{ds}$ & \\
 $\Checked{oks}{fs} \oplus \Validate$ & = $\Checked{oks}{fs}$ & \\
 $\Checked{oks}{fs} \oplus \Checked{oks'}{fs'}$ & = $\Checked{oks\cup{}oks'}{fs\cup{}fs'}$ & \\
 $\Checked{oks}{fs} \oplus \WaitFor{ds}$ & = $\Checked{oks\cup{}ds}{fs\cup{}fs}$ & \\
 $\WaitFor{ds} \oplus \Validate$ & = $\WaitFor{ds}$ & \\
 $\WaitFor{ds} \oplus \Checked{oks}{fs}$ & = $\Checked{oks\cup{}ds}{fs}$ & \\
 $\WaitFor{ds} \oplus \WaitFor{ds'}$ & = $\WaitFor{ds\cup{}ds'}$ & \\
\end{tabular}

The algorithm presented in figure~\ref{PSchema} required as parameters a function \c|checkLocal|: (\LabelSet, \VertSet) $\rightarrow$ \Ok $\mid$ \Failed $\mid$ \Pending(\c|Set|[\LabelSet]) that returns \Ok\ if it is possible to check that the node conforms to a shape label locally, \Failed\ if it is possible to check that a node doesn't conform to a shape label locally, and \Pending(ls)\ if the conformance of a node depends on the arcs in $ls$. 

Figure~\ref{alg:checkLocal} presents a possible implementation of \c|checkLocal| for WShEx. 

\begin{algorithm}[!hbt]
  \DontPrintSemicolon
  \SetAlgoVlined
  \DefInline{\checkLocal{\lbl}{n}}{\checkLocal{\schemaDef(\lbl)}{n}}  \\
  \DefnCustom{\checkLocal{se}{n}=\Match~se}{
     \Case~~$se_1$ \c|AND| $se_2$\MapTo$\combine{\checkLocal{se_1}{n}}{\checkLocal{se_2}{n}}$\\
     \Case~~@\lbl \MapTo $\checkLocal{\schemaDef(\lbl)}{n}$\\
     \Case~~\{ $te$ \} \MapTo $\checkLocal{te}{n}$ \\
     \Case~~\c|CLOSED| \{ $te$ \} \MapTo $\checkLocal{te}{n}$ \\
    }
    \DefnCustom{$\checkLocal{te}{n}=\Match~te$}{
     \Case~~$\eachOf{te_1}{te_2}$\MapTo$\combine{\checkLocal{te_1}{n}}{\checkLocal{te_2}{n}}$ \\
     \Case~~$\someOf{te_1}{te_2}$\MapTo$\combine{\checkLocal{te_1}{n}}{\checkLocal{te_2}{n}}$ \\
     \Case~~$te*$\MapTo
       $\checkLocal{te}$ \\
     \Case~~\arc{p}{@\lbl\,\,qs}\MapTo
       $\Pending{\{\lbl\}}$
    }
    \DefnCustom{$\combine{r_1}{r_2}=\Match~(r_1,r_2)$}{
     \Case~~(\Ok,\Ok)\MapTo\Ok \\
     \Case~~(\Ok,\Pending{ls})\MapTo\Pending{ls} \\
     \Case~~(\Ok,\Failed)\MapTo\Failed \\
     \Case~~(\Pending(ls),\Ok)\MapTo\Pending{ls} \\
     \Case~~(\Pending($ls_1$),\Pending($ls_2$))\MapTo\Pending($ls_1\cup{}ls_2$) \\
     \Case~~(\Pending(ls),\Failed)\MapTo\Failed \\
     \Case~~(\Failed,\_)\MapTo\Failed \\
    }
  \caption{Definition of \c|checkLocal|:(\LabelSet, \VertSet) $\rightarrow\Ok \mid \Failed \mid \Pending(\c|Set|[\LabelSet])$}
  \label{alg:checkLocal}
\end{algorithm}

The definition of \c|checkNeighs|: (\LabelSet, \c|Bag|[(\EdgeSet, \LabelSet)],  \c|Set|[(\EdgeSet, \LabelSet)]) $\rightarrow$ \Ok$\mid$\Failed\  is shown in figure~\ref{alg:checkNeighs}.

\begin{algorithm}[!hbt]
  \DontPrintSemicolon
  \SetAlgoVlined
  \DefInline{\checkNeighs{\lbl}{\bagw}{fs}}{\checkNeighs{\schemaDef(\lbl)}{\bagw}{fs}}  \\
  \DefnCustom{$\checkNeighs{se}{\bagw}{fs}$=\Match~se}{
     \Case~~$cond$\MapTo$\Ok$ \\ 
     \Case~~$se_1$ \c|AND| $se_2$\MapTo$\checkNeighs{se_1}{\bagw}{fs}\wedge{}\checkNeighs{se_2}{\bagw}{fs}$\\
     \Case~~@\lbl \MapTo $\checkNeighs{\schemaDef(\lbl)}{\bagw}{fs}$\\
     \Case~~\{ $te$ \} \MapTo $\matchRbe{\bagw}{\rbe{te}}$ \\
     \Case~~\c|CLOSED| \{ $te$ \} \MapTo $\matchRbe{\bagw}{\rbe{te}}\wedge{}fs=\emptyset$
    }
  \DefnCustom{$\rbe{te}$=\Match~te}{
     \Case~~\eachOf{te_1}{te_2} \MapTo$\rbe{te_1}; \rbe{te_2}$\\
     \Case~~\oneOf{te_1}{te_2} \MapTo $\rbe{te_1} \mid \rbe{te_2}$\\
     \Case~~$te*$ \MapTo $\rbe{te}*$ \\
     \Case~~\arc{p}{@\lbl\,\,qs}\MapTo $(p,\lbl)$
    }
  \caption{Definition of \c|checkNeighs|: (\LabelSet, \c|Bag|[(\EdgeSet, \LabelSet)],  \c|Set|[(\EdgeSet, \LabelSet)]) $\rightarrow\Ok\mid\Failed$}
  \label{alg:checkNeighs}
\end{algorithm}
 
Finally, \c|tripleConstraints|: \LabelSet $\rightarrow$ \c|Set|[(\EdgeSet, \LabelSet)]| returns the triple constraints associated with a shape label. 
It is defined in figure~\ref{alg:tripleConstraints}

\begin{algorithm}[!hbt]
  \DontPrintSemicolon
  \SetAlgoVlined
  \DefnCustom{$\tripleConstraints{se}$=\Match~se}{
     \Case~~$cond$\MapTo\{\} \\
     \Case~~$se_1$ \c|AND| $se_2$\MapTo$\tripleConstraints{se_1}\cup\tripleConstraints{se_2}$\\
     \Case~~@\lbl \MapTo \c|tripleConstraints|($\delta(l)$) \\
     \Case~~\{ $te$ \} \MapTo $\tripleConstraints{te}$ \\
     \Case~~\c|CLOSED| \{ $te$ \} \MapTo $\tripleConstraints{te}$
    }
 \DefnCustom{$\tripleConstraints{te}$=\Match~te}{
     \Case~~\eachOf{te_1}{te_2} \MapTo$\tripleConstraints{te_1}\cup\tripleConstraints{te_2}$\\
     \Case~~\oneOf{te_1}{te_2} \MapTo $\tripleConstraints{te_1} \cup \tripleConstraints{te_2}$\\
     \Case~~$te*$ \MapTo $\tripleConstraints{te}$ \\
     \Case~~\arc{p}{@\lbl\,\,qs}\MapTo $\{(p,\lbl)\}$
    }
  \caption{Definition of \c|tripleConstraints|: \LabelSet $\rightarrow$ \c|Set|[(\EdgeSet, \LabelSet)] }
  \label{alg:tripleConstraints}
\end{algorithm}

\begin{example}[Pregel+ShEx example]

As an example, we will use the Wikibase graph from example~\ref{example:WikibaseGraph} to validate the ShEx schema from example~\ref{example:ShExSlurp}. We replace the shape labels by their initial so we will use:

\begin{tabular}{cl}
\LabelSet & = \{ $Researcher$, $Place$, $Country$, $Date$, $Human$\} \\
\schemaDef($R$) & = \{ \arc{\instanceOf}{@$H$}; \arc{\birthDate}{@$D$}; \arc{\birthPlace}{@$P$} \} \\
\schemaDef($P$)  & = \{ \arc{\country}{@$C$}\} \\
\schemaDef($C$)  & = \{ \} \\
\schemaDef($D$)  & = $\in{}xsd:date$\\
\schemaDef($H$)  & = $\in{}\{\Human\}$\\
\end{tabular}

The first step of the algorithm will send a message to every node requesting it to validate with the shape $Researcher$, 
 and after running $vProg$ the status of every node will be $\Pending$ on shape $Researcher$. 
In the first superstep, the messages that will be generated by each triple are\footnote{For simplicity, for each node $(x,m_x)$ we show only $x$ and we omit qualifiers in the triples as they are always empty}:

\begin{tabular}{l|l}
Triple & Messages \\ \hline
\multirow{2}{*}{(\timBl, \birthPlace, \London)} & $\msgSent{\timBl}{R}{\WaitFor{(\London,\birthPlace,P)}}$ \\
           & $\msgSent{\London}{P}{\Validate}$ \\ \hline
\multirow{2}{*}{(\timBl, \instanceOf, \Human)} & $\msgSent{\timBl}{H}{\WaitFor{(\Human,\instanceOf,H)}}$ \\
           & $\msgSent{\Human}{H}{\Validate}$ \\ \hline
\multirow{2}{*}{(\vintCerf, \birthPlace, \newHaven)} & $\msgSent{\vintCerf}{P}{\WaitFor{(\newHaven,\birthPlace,P)}}$ \\
           & $\msgSent{\Human}{H}{\Validate}$ \\ \hline
\multirow{2}{*}{(\vintCerf, \instanceOf, \Human)} & $\msgSent{\vintCerf}{H}{\WaitFor{(\Human,\instanceOf,H)}}$ \\
           & $\msgSent{\Human}{H}{\Validate}$ \\ \hline
\end{tabular} 

After running $vProg$ the status of all nodes except \timBl and \vintCerf with regards to the label $R$ will be $\Failed$ because they will fail to $checkLocal$. The status of both \timBl and \vintCerf will be waiting for the validation of their neighborhood nodes.

After superstep 2, the messages generated will be:

\begin{tabular}{l|l}
Triple & Messages \\ \hline
\multirow{2}{*}{(\London, \country, \UK)} & $\msgSent{\London}{P}{\WaitFor{(\UK,\country,C)}}$ \\
           & $\msgSent{\UK}{C}{\Validate}$ \\ \hline
(\timBl, \instanceOf, \Human) & $\msgSent{\timBl}{R}{\Checked{(\Human,\instanceOf,H)}{\{\}}}$ \\ \hline
(\vintCerf, \instanceOf, \Human) & $\msgSent{\vintCerf}{P}{\Checked{(\Human,\instanceOf,H)}{\{\}}}$ \\ \hline
(\vintCerf, \birthPlace, \newHaven) & $\msgSent{\vintCerf}{P}{\Checked{(\{\})}{(\newHaven,\birthPlace,P)}}$ \\
\end{tabular} 

After running $vProg$, the status of \UK will be \Ok for shape label $C$, the status of \London will be waiting for \UK to validate as $C$, the status of \vintCerf will be \Failed for shape label $R$ (it fails because the value \birthPlace failed). 
In the third superstep, the messages generated will be:

\begin{tabular}{l|l}
Triple & Messages \\ \hline
(\London, \country, \UK) & $\msgSent{\London}{P}{\Checked{(\UK,\country,C)}{\{\}}}$ \\ \hline
\end{tabular} 

Which will change the status of \London to \Ok for shape label $P$. In the fourth superstep, the messages that will be sent are:

\begin{tabular}{l|l}
Triple & Messages \\ \hline
(\timBl, \birthPlace, \London) & $\msgSent{\timBl}{R}{\Checked{(\London,\birthPlace,P)}{\{\}}}$ \\ \hline
\end{tabular} 

And after running $vProg$ the status of \timBl for shape label $R$ will be \Ok. 
In the next superstep, no more messages will be sent and all the nodes will be inactive, finalizing the Pregel algorithm. 
 The nodes that have a \Ok status for some shape are:
 
\begin{tabular}{l|l}
Node & Shape Label \\ \hline
\timBl & R \\
\London & P \\
\UK & C \\
\end{tabular} 

And the generated subset will consist of collecting information about those nodes:

\begin{tabular}{ccl}
$\rho$ &= \{ & (\timBl, \instanceOf, \Human, \{\}), \\ 
 & & (\timBl, \birthDate, 1955, \{\}), \\ 
 & & (\timBl, \birthPlace, \London, \{\}), \\
 & & (\London, \country, \UK, \{\}), \\
\end{tabular}
\end{example} 

\paragraph{Representing Wikidata in Spark GraphX}

Spark GraphX supports a kind of property graphs where vertices and edges can have an associated value. 
 The type \lstinline|Graph[VD,ED]| is used to represent a graph whose vertices are pairs of values of type  
\c|(VertexId,VD)| where \c|VertexId=Long| represents a unique vertex identifier and \c|VD| represents the value associated with the vertex.

When representing Wikidata graphs in Spark GraphX it is necessary to take into account which entities will be represented as vertices. 
 We opted to include as nodes in the graph only those nodes that can be subjects of statements, i.e. wikidata entities (items and properties), 
  leaving out primitive values like literals, dates, etc. 
  We separated the statements associated with an entity in two kind of statements: local statements, which are embedded inside the value of a node and whose values can be accessed 
   without traversing the graph, and entity statements, whose values are other entities which have their corresponding vertex in the graph.

In this way, the WShEx also needs to distinguish triple expressions between local ones and entity ones.   

%
%

%% file: 60_Results.tex
\section{Implementations and preliminary results} \label{sec:Results}

Although we have implemented some of the approaches presented in the paper, 
 further work needs to be done in order to have all the possible implementations. 
 In table~\ref{table:Implementations}, we present an overview of the approaches presented in the paper and their implementation. 
 
\newcommand{\pending}{--}
\newcommand{\thispaper}{[This paper]}

\index{WDumper} \index{PyShEx} \index{shex.js} 
 \begin{table}[h!]
 \begin{tabular}{m{12em}m{7em}m{7em}m{7em}}
  & RDF graphs & \emph{Property graphs} & \emph{Wikibase graphs} \\ \hline
  Formal definition & Done, ex.\cite{Gutierrez03} & Done, \cite{Prudhommeaux2014, Labra-Gayo2019} & \thispaper \\ \hline
  Description \newline with Shape Expressions & ShEx\newline\cite{Prudhommeaux2014, Slawek2015, Boneva17} & PShEx\newline\thispaper & WShEx\newline\thispaper \\ \hline
  Entity-generated subsets & \pending & \pending & \href{https://github.com/bennofs/wdumper}{WDumper} \\ \hline
  Matching generated subsets & \pending & \pending & \href{https://github.com/bennofs/wdumper}{WDumper} \\ \hline
  ShEx-based matching & \pending & \pending & \href{https://github.com/weso/wdsub}{WDSub} \\ \hline
  ShEx+Slurping & \href{https://github.com/shexjs/shex.js}{shex.js} \newline{} \href{https://github.com/hsolbrig/PyShEx}{PyShEx} & \pending & \pending \\ \hline
  ShEx + Pregel & \pending & \pending & \href{https://github.com/weso/sparkwdsub}{SparkWDSub} \\ \hline
 \end{tabular}
 \caption{Overview of implementation/formal definition status} \label{table:Implementations}
 \end{table}

The two implementations created by the authors of this paper are~\href{https://github.com/weso/wdsub}{WDSub} and~\href{https://github.com/weso/sparkwdsub}{SparkWDSub}.
 They have been implemented in Scala and are mainly proof-of-concepts 
  which have not yet been thoroughly tested and optimized 
  so the results presented here are preliminary. 

\index{WDSub}
WDSub takes as input a simple ShEx file containing the schema and another 
 file with a Wikibase dump. It extracts the subset that matches the ShEx without taking into account shape references. It uses two approaches, one based on 
 Wikidata toolkit\footnote{\url{{https://www.mediawiki.org/wiki/Wikidata_Toolkit}}}, 
 and another one based on the fs2 library\footnote{\url{https://fs2.io/}}. 
 We tested it using a virtual server running on XCP-ng 8.2 (CentOS 8 stream) with 32 cores and 64Gb of RAM and it was able to process the JSON Wikidata dump 
 from 2021, which has 1.256,55Gb uncompressed, in 5h 15min.

SparkWDSub implements the Pregel+ShEx approach. It has been built on top of the Spark GraphX library~\footnote{\url{http://spark.apache.org/graphx/}}. 
 In order to obtain greater flexibility, it defines a generic \c|PSchema| class\footnote{\url{https://github.com/weso/sparkwdsub/blob/master/src/main/scala/es/weso/pschema/PSchema.scala}} which defines a generic graph validation algorithm. 

\lstset{language=scala}

Given a target graph \c|Graph[VD,ED]| of vertex \c|VD| and edges \c|ED|, after running Pregel, it will return \c|ShapedGraph[VD,L,E,P]| 
 which will contain information about the shapes labels \c|L| associated with a vertex \c|VD| in a graph with properties \c|P| 
 and possible errors \c|E| and the \c|MsgMap[L,E,P]| class that represents the messages sent to a vertex.
 
 The \c|PSChema| class is defined as follows:
 
\begin{lstlisting}[language=scala]
class PSchema[VD, ED, L, E, P]
     (checkLocal: (L, VD) => Either[E, Set[L]], 
      checkNeighs: (L, Bag[(P,L)], Set[(P,L)]) => Either[E, Unit],
      getTripleConstraints: L => List[(P,L)], 
      cnvEdge: ED => P,
     ) {
  def vprog(id: VertexId, 
            g: Shaped[VD, L, E, P], 
            msg: MsgMap[L, E, P]): Shaped[VD, L, E, P] = { . . . }
  
  def sendMsg(t: EdgeTriplet[Shaped[VD, L, E, P],ED]
             ): Iterator[(VertexId, MsgMap[L, E, P])] = { . . . }

  def mergeMsg(p1: MsgMap[L,E,P], 
               p2: MsgMap[L,E,P]): MsgMap[L,E,P] = { . . .}
  
  def runPregel(graph: Graph[VD,ED],
                initialLabel: L,
                maxIterations: Int = Int.MaxValue,
               ): Graph[Shaped[VD,L,E,P],ED] = {
    Pregel(shapedGraph(graph),
           initialMsg(initialLabel),
           maxIterations)(vprog, sendMsg, mergeMsg)
    .mapVertices(checkRemaining)
  }
}
\end{lstlisting}

\noindent{}where \c|checkRemaining| checks which of the nodes with \c|Pending| status are valid or not.

The previous interface enables the possible implementation with different validation languages like ShEx, PShEx or WShEx.  

\index{SparkWDSub}
SparkWDSub currently offers an implementation for a simplified version of WShEx which implements the \c|checkLocal|, \c|checkNeighs|, \c|getTripleConstraints| and \c|cnvEdge| to convert an edge \c|ED| into a property \c|P|.

We have run an experiment on AWS using 512 cores, 3.904 Gb, 121.600 Gb, 
and it was possible to generate subsets for the 2014 Wikidata dump (31.3 GB) in 3 minutes, 
while with the 2021 Wikidata dump (1.256,55 Gb uncompressed) it took 36 minutes.

In the experiments we used the following ShEx schema to obtain cities:

\begin{lstlisting}[style=ShExC]
prefix wde: <http://www.wikidata.org/entity/>

Start = @<City>

<City> {
 wde:P31 @<CityCode> 
}
<CityCode> [ wde:Q515 ]
\end{lstlisting}

%% file: 70_RelatedWork.tex
\section{Related work} \label{sec:RelatedWork}

\index{Knowledge graphs}
\paragraph{Knowledge graphs}~\\
An introduction to Knowledge graphs is provided by~\cite{Hogan2021}, 
which cites other 
books~\cite{PVGW2017, Kejriwal19, FenselSAHKPTUW20} 
and surveys like~\cite{Paulheim17,WangY19,YanWCGZ18,GeseseBS19,XiaoDCC19,Al-MoslmiOOV20}.
In this paper, we follow two of the graph models provided in that survey: directed labeled graphs, which we call RDF-graphs, 
and property graphs; and add a new one: wikibase graphs. 
\index{MARS} \index{Multi-Attributed Relational Structures} \index{MAPL} \index{Multi-Attributed Predicate Logic}
 Our definition of wikibase graphs has been inspired by MARS (Multi-Attributed Relational Structures)~\cite{MKT2017}, which are a 
  a generalized notion of property graphs. 
  In that paper, they also define MAPL (Multi-Attributed Predicate Logic) as a logical formalism that can be used for ontological reasoning.


\index{ShEx} \index{SHACL} \index{W3C recommendation}
\paragraph{Knowledge graph descriptions}~\\
Since the appearance of ShEx in 2014, there has been a lot of interest about RDF validation and description. 
In 2017, the data shapes working group proposed SHACL (Shapes Constraint Language) as a W3C recommendation~\cite{SHACLSpec}. 
Although SHACL can be used to describe RDF, its main purpose is to validate and check constraints about RDF data makes it less usable to describe RDF subsets.

\index{Entity schemas}
ShEx was adopted by Wikidata in 2019 to define entity schemas~\cite{ThorntonSSGMPW19}. We consider that ShEx adapts better to describe data models than SHACL, which is more focused on constraint violations. 
A comparison between both is provided in~\cite{Labra2017} while in~\cite{Labra-Gayo2019}, a simple language is defined 
that can be used as a common subset of both.

Improving quality of Knowledge graphs in general, and Wikidata in particular, has been the focus of some recent research like~\cite{Piscopo2019,Turki2020,Shenoy21}.

\index{SQID}
Following the work on MARS, there has been some recent work about adding an inference layer on top of Wikidata.
 The project SQID~\cite{MK2017} combines inference and visualization to create a Wikidata browser. 
 Another possibility that has been explored is to use MARS reasoning to define constraints~\cite{Martin2020}.

\paragraph{Big data processing and graphs}~\\

\index{Pregel} \index{GraphLab} \index{PowerGraph} \index{GraphX}
There has been a lot of interest in the last decade to develop scalable algorithms that can process big data graphs. 
 In 2010, Pregel was proposed by Google~\cite{Malewicz2010} as a suitable model for large-scale graph computing.
 Following that publication, several systems were developed like GraphLab~\cite{Low2012}, PowerGraph~\cite{PowerGraph12}
 and GraphX~\cite{GraphX14} which followed the lemma \emph{thinking like a vertex}\cite{McCune2015} where scalable graph are based on local iterations over the nodes of a graph and their neighbors. 

\index{RDD} \index{Resilient Distributed Datasets} 
 GraphX was a framework that internally represented graphs using Apache Spark's Resilient Distributed Datasets (RDDs)~\cite{Zaharia12} 
  enabling to implement graph-parallel abstractions and algorithms like Pregel.
A functional definition of Spark using Haskell has been proposed in~\cite{Chen2017}. 
 It would be interesting to use that functional specification to prove the correctness of the algorithm proposed in this paper.
Another related line of work is the creation of domain specific languages to facilitate encoding of Pregel-based algorithms like Palgol~\cite{Zhang2017} or Fregel~\cite{Emoto2016}.  

\paragraph{Knowledge graphs subsets}

\index{RDF} \index{SPARQL}
Although it is possible to create subsets of RDF graphs using SPARQL construct queries, the approach usually requires some scripts 
 to launch the queries as SPARQL doesn't support recursion so it is not possible to represent cyclic data models. 
 There has been a proposal to extend SPARQL with recursion~\cite{Reutter2015}, but it is not part of 
  most existing SPARQL processors.

\index{RDFSlice}
RDFSlice~\cite{RDFSlice2013, Marx2013} was proposed as a system that generated RDF data fragments from
 large endpoints like \indexn{DBpedia}. 
  It defines a subset of SPARQL called \indexn{sliceSPARQL}. 
  A new version, called \indexn{Torpedo}, was proposed in~\cite{Torpedo2017}, which improves the perfomance and adds further expressivity. 
  The use of SPARQL to generate subsets is one important difference with the work presented in this paper. 
  We consider that ShEx improves the expressiveness and usability of SPARQL to describe data models and subsets as it allows 
   cyclic or recursive data models and has a declarative syntax that is specifically defined fur such endeavor.
 
\index{SWAT4HCLS}  \index{Wikidata} \index{Biohackathon}
The creation of Wikidata subsets has been a topic of interest since the 12th International SWAT4HCLS Conference\footnote{This page was created to collect information: \url{https://www.wikidata.org/wiki/Wikidata:WikiProject_Schemas/Subsetting}}. 
 It was later selected as a topic in the Elixir Europe Biohackathon 2020\footnote{\url{https://github.com/elixir-europe/BioHackathon-projects-2020/tree/master/projects/35}} and the 
 SWAT4HCLS 2021 hackathon from which a prepring was generated that collects the different approaches proposed~\cite{LabraWikidataSubsetting21}.

\index{WDumper}
One of the approaches was 
WDumper~\footnote{\url{https://github.com/bennofs/wdumper}}, which was created as a tool that processed Wikidata dumps and generates subsets from those dumps. 
It takes two inputs: a JSON compressed dump and a JSON configuration file that describes the different filters,  
 and generates as output an RDF compressed dump.
 The tool can be run as a web service locally and is also deployed at~\url{https://tools.wmflabs.org/wdumps}. 
 It also contains a web service allows the user to introduce the different filters filling a form which generates a dump generation request 
  that is added to a queue.
 It is possible to see the list of previously requested dumps. Once the dump has been generated, 
  it can be uploaded to \indexn{Zenodo}.  
 WDumper is divided in two main modules, the backend, which has been implemented in Java using the Wikidata Toolkit library~\footnote{\url{https://github.com/Wikidata/Wikidata-Toolkit}} 
 and the frontend that has been implemented in Typescript. 
 The use of WDumper to generate Wikidata subsets is described in~\cite{Seyed21} where 4 subsets are created about the topics:
politicians, 
military politicians, 
UK universities and 
GeneWiki data. That paper also presents several use case scenarios and 
 discusses some strengths and weaknesses.
In this paper, we present a formal definition of WDumper in the context of other subset generation approaches 
 like the ShEx-based ones. 




The Python library WikidataSets~\cite{Wikidatasets2019} generates Wikidata subsets from specific topics. 
 In the paper the authors generated subsets for the following topics: humans, countries, companies, animal species and films.
 The tool obtains items following the instances of a topic or subclasses of the topic.

\index{KGTK} \index{Kypher} \index{SQLite}
KGTK (Knowledge graph toolkit)~\cite{Ilievski2020} is a tool that works with 
knowledge graphs by defining a common format called KGTK format based on hypergraphs. 
It is possible to import and export data from different formats like \indexn{Wikidata} or \indexn{ConceptNet}, 
do several operations over those graphs like: validation, cleaning, graph manipulation 
(sort, column removal, edge filtering) 
and graph merging (join, cat) operations. 
The tool also supports graph querying and analytics operations. 
Given that KGTK can take as input Wikidata dumps and generate Wikidata dumps as output, 
 it is possible to use KGTK to generate subsets of Wikidata. 
More recently, the authors have published a paper where they apply KGTK to create personalized versions of wikidata using a query language that they call Kypher\cite{Chalupsky21}, which is an adapted version of Cypher for KGTK.
 Internally, it uses a tabular data model which allows to translate Kypher queries to SQLite\footnote{\url{https://www.sqlite.org/}}. 
 According to the paper, it can create very efficient queries processing the whole Wikidata in a laptop system. 
 The main difference with our approach is that we use Shape Expressions to describe the data model of the subset. 
 Further work could be done to see if it was possible to translate Shape Expression definitions to Kypher.

%% file: 80_conclusions.tex
\section{Conclusions} \label{sec:Conclusions}

\index{RDF-based graphs} \index{Property graphs} \index{Wikibase graphs}
In this paper, we have presented three formal models for knowledge graphs: RDF-based graphs, property graphs and wikibase graphs. 
\index{WShEx} \index{PShEx} \index{ShEx}
 We also defined a shape expressions language that can be used to describe and validate data in those models: ShEx for RDF-based graphs, PShEx for property graphs and WShEx for wikibase graphs.

Given the success of knowledge graphs, their size has been increasing in a way that it is not possible to process their contents using conventional tools making it necessary 
 to have some mechanism to extract subsets from them. 
 Finally, we review some approaches to generate subsets from Wikibase graphs. 
  The first two approaches, entity-matching and simple matching can be implemented by processing Wikibase dumps sequentially. 
  The third approach takes as input a WShEx schema, and matches the different entities and their local neighbors with the shapes ignoring shape references.
   This approach can be used to efficiently process dumps sequentially but doesn't take into account the relations in the graph.
  \index{ShEx+Slurp} 
  The fourth approach, ShEx+Slurp, adds an option to the ShEx processor to collect the visited triples while it is validating the data. 
   This approach can do graph traversal but also require a large number of requests for nodes neighbors which may not be possible to apply it behind endpoints that limit the number of requests. 
  \index{ShEx+Pregel} 
   The final approach that we proposed applies the Pregel algorithm to validate all nodes.
    This approach does graph traversal and can also handle large graphs. 

 All the approaches have been implemented, although not all of them within the same system, so a proper comparison is not yet possible.  
 Further work needs to be done on improving the implementations, applying them to some use cases and assessing their advantages and challenges.

%% file: 100_acknowledgments.tex
\section{Acknowledgments}

We want to thank Eric Prud'hommeaux, Andra Waagmeester, Tom Baker, Kat Thornton and Iovka Boneva as well as all the members of the Shape Expressions Community Group. 
The Wikidata and Wikibase community provided a great environment and software whose potential we are still discovering, special thanks to Lydia Pintscher,  Adam Shorland, Daniel Mietchen, Egon Willighagen, Finn {\AA}rup Nielsen, etc.
Some parts of this research took inspiration from several events like 2019 and 2020 biohackathons and SWAT4HCLS. 
Special thanks to Dan Brickley for helping with the coordination and insights on the Wikidata subsetting project.
Markus Krötzsch's lectures and papers have also been a source of inspiration for the Wikibase formal model and Tim Berners-lee's running example.

Guillermo Facundo Colunga, Pablo Menéndez Suárez, Jorge Álvarez Fidalgo and Daniel Fernández Álvarez from the WESO research group have also helped with the experiments on Wikidata Subsetting and Scholia.